\documentclass[twocolumn]{aastex62}

\usepackage{amsmath}
\usepackage{longtable}
\usepackage{natbib}
 \usepackage{appendix}
\usepackage{array,multirow}
\usepackage{nicematrix}
\usepackage{lineno}

\shorttitle{Vertical Structure of WASP-76b}
\shortauthors{Kesseli \& Beltz et al.}
\pdfoutput=1

\begin{document}

\title{Up, Up, and Away: Winds and Dynamical Structure as a Function of Altitude in the Ultra-Hot Jupiter WASP-76b}

\correspondingauthor{Aurora Y. Kesseli}
\email{aurorak@ipac.caltech.edu}
\author[0000-0002-3239-5989]{Aurora Y. Kesseli}
\altaffiliation{These authors contributed equally to this work.}
\affiliation{IPAC, Mail Code 100-22, Caltech, 1200 E. California Blvd., Pasadena, CA 91125, USA}

\author[0000-0002-6980-052X]{Hayley Beltz}
\altaffiliation{These authors contributed equally to this work.}
\affiliation{Department of Astronomy, University of Maryland, College Park, MD 20742, USA}

\author[0000-0003-3963-9672]{Emily Rauscher}
\affiliation{Department of Astronomy, University of Michigan, Ann Arbor, MI 48109, USA}

\author[0000-0003-1624-3667]{I.A.G. Snellen}
\affiliation{Leiden Observatory, Leiden University, Postbus 9513, 2300 RA, Leiden, The Netherlands}

\begin{abstract}

Due to the unprecedented signal strengths offered by the newest high-resolution spectrographs on 10-m class telescopes, exploring the 3D nature of exoplanets is possible with an unprecedented level of precision.
In this paper, we present a new technique to probe the vertical structure of exoplanetary winds and dynamics using ensembles of planet absorption lines of varying opacity, and apply it to the well-studied ultra-hot Jupiter WASP-76b. We then compare these results to state-of-the-art global circulation models (GCMs) with varying magnetic drag prescriptions. We find that the known asymmetric velocity shift in Fe I absorption during transit persists at all altitudes, and observe tentative trends for stronger blueshifts and more narrow line profiles deeper in the atmosphere.  
By comparing three different model prescriptions (a hydrodynamical model with no drag, a magnetic drag model, and a uniform drag model) we are able to rule out the uniform drag model due to inconsistencies with observed trends in the data. 
We find that the magnetic model is slightly favored over the the hydrodynamic model, and note that this 3-Gauss kinematic magnetohydrodynamical GCM is also favored when compared to low-resolution data.
Future generation high-resolution spectrographs on Extremely large telescopes (ELTs) will greatly increase signals and make methods like these possible with higher precision and for a wider range of objects.

\vspace{25pt}

\end{abstract}

\section{Introduction}

Transmission spectroscopy has dominated the study of exoplanet atmospheres since the first atmospheric detection \citep{Charbonneau2002}, and has moved the study of exoplanets from bulk properties to in-depth studies of composition, chemical processes, and dynamics in exoplanet atmospheres. Transmission spectroscopy performed at high spectral resolution has proven to be a powerful technique for detecting a range of atomic and molecular species that are often difficult to distinguish at lower spectral resolution \citep[e.g.,][]{Hoeijmakers2019, Giacobbe2021, Kesseli2022}. By taking advantage of the fine radial velocity spacing of the spectrographs, high resolution spectroscopy has also been a valuable tool to uncover winds and atmospheric dynamics \citep[e.g.,][]{Snellen2010}.

Hot and ultra-hot Jupiters offer the best test beds for assessing both chemical models and dynamical global circulation models (GCMs) for exoplanetary atmospheres due to their large scale-heights and favorable star-to-planet radius ratios. Recent instrumentation advances from facilities such as \textit{JWST}, and increased throughput and wavelength coverage of next generation high-resolution spectrographs like ESPRESSO \citep{Pepe2014}, MAROON-X \citep{seifahrt2018}, KPF \citep{Gibson2016}, and IGRINS \citep{park14}, mean that more complex planetary models can be tested. GCMs are 3D numerical models that simulate the dynamics of planetary atmospheres. These models are computationally expensive, but can capture many of the complex 3D processes such as clouds \citep{Harada2021, Malsky2021, savel2022}, scale height effects \citep{Savel2023,Wardenier2023}, magnetic drag \citep{RauscherMenou2013,Beltz2022a}, hydrogen dissociation \citep{TanKomacek2019}, and changing line contrasts due to a 3D temperature structure \citep{Lee2022, vansluijs2023} that influence the atmosphere and resulting spectra. 

The publicly available ESPRESSO dataset consisting of two transits of WASP-76b represents the highest-signal-to-noise-ratio high-resolution transmission spectrum that is publicly available to date. This dataset has been used in many papers due to its unprecedented quality, including the novel detection of asymmetric Fe I absorption during transit \citep{Ehrenreich2020}, and subsequent detection and analysis of asymmetries in many other species \citep{Kesseli2022}. It was also used to detect single absorption lines for a range of metals \citep[i.e., Na, Li, Mg, Fe, K; ][]{Tabernero2021, Seidel2021}, and in retrievals to produce phase- and spatially-resolved  parameters (abundances, temperature-pressure profiles) for the first time \citep{Gandhi2022}. Most recently, \citet{Gandhi2023} also precisely constrained the abundances of eight different atomic species.

In this paper, we use the same WASP-76~b data and present a novel method to probe 3D atmospheric dynamics by exploring how radial velocities and line shapes change as a function of altitude (vertically). As was shown by \citet{Kempton2012}, measurements of how wind speeds and net Doppler shifts vary in altitude may help differentiate between different physical conditions input into GCMs, such as magnetic field strength and drag parameters. We aim to use the added vertical dimensionality offered by this technique to put more robust constraints on the physical processes (e.g., drag, magnetism, winds) occurring within WASP-76~b. 

We then compare these results to output from our state-of-the-art GCMs. This paper focuses on the treatment of drag in GCM models as the drag prescription is the main process that sets the wind speeds, and in turn the main influencer on measured radial velocities. Here we use the RM-GCM\footnote{https://github.com/emily-rauscher/RM-GCM} which was the first GCM to be post-processed and directly compared to high resolution emission \citep{Beltz2021} and transmission \citep{Flowers2019} spectra. These works, among others \citep{Wardenier2021, Beltz2022b, Gandhi2022, Beltz2023, Pluriel2023, Savel2023, vansluijs2023, Wardenier2023} highlight the inherent ``3Dness'' present in high resolution spectroscopic observations. An extremely multidimensional aspect of our models is the inclusion of a spatially varying magnetic drag. This state-of-the-art active magnetic drag prescription allows for the strength of magnetic drag to be calculated based on local conditions. Most GCMs use a single ``uniform'' drag timescale to simulate the effects of magnetic drag \citep{KomacekShowman2016, Kreidberg2018, Mansfield2018, Arcangeli2019,Lee2022}. This assumption is particularly problematic for ultra-hot Jupiters (UHJs), as it can imply unphysically large magnetic field strengths on the nightside of the planet \citep[see discussion section in ][for an order of magnitude calculation]{Beltz2022a}. Previous works have shown the inclusion of magnetic effects can alter predicted phase curves by reducing hotspot offsets and increasing amplitude due to an increased day-night contrast \citep{Beltz2022a}. This is in agreement with observations of ultra-hot Jupiters, which often show smaller hotspot offsets than their hot Jupiter counterparts \citep{ Zhang2018, TanKomacek2019, May2022}. Additionally, the inclusion of magnetic drag can result in net Doppler shifts that can vary on the order of several kilometers per second in high resolution emission \citep{Beltz2022b} and transmission \citep{Beltz2023} spectra. We conduct our analysis on a uniform drag as well as a drag-free model for comparison to the active drag case. 

We present the data in Section \ref{s:data}, and our cross correlation analysis in Section \ref{s:cc}. The GCM models are then described in Section \ref{s:gcm}. We present the results of the cross correlation analysis on the WASP-76~b ESPRESSO data in Section \ref{s:results_fe} and \ref{s:results_other}, and compare the observed trends in altitude present in the data to state-of-the-art GCM models in Section \ref{s:comp_gcm}. We discuss how our results fit in with previous work and potential future directions for GCMs in Section \ref{s:disc}. Finally, we summarize our conclusions in Section \ref{s:conc}. 

\section{Observations and Data Analysis} 
\label{s:data}

We downloaded the 1D blaze-corrected, stitched spectra from the Data and Analysis Center for Exoplanets (DACE) database\footnote{\url{https://dace.unige.ch/dashboard/index.html}}, which were reduced with version 1.3.2 of the ESPRESSO pipeline. More details on the observing strategy are given in \citet{Ehrenreich2020}. Before any cleaning or analysis steps, we removed contamination by telluric H$_2$O and O$_2$ lines using molecfit \citep{smette2015}, which has proven adept at removing the shallow absorption lines that mainly permeate the optical portion of the spectrum \citep[e.g.,][]{allart2017}. 

We outline the general cleaning steps we performed here, but refer the reader to \citet{Kesseli2022} for more details, as the data presented here are the same as those used in that study. In order to remove the contribution from the host star so that the much weaker signal from the exoplanet can be seen, we created a uniform spectral time series along both the wavelength and flux axes. We performed the cleaning steps on each of the two nights separately, and then combined the nights after cross correlation. Using the host star's known radial velocity induced by the planet \citep[$K_{*}=0.1156$ km s$^{-1}$;][]{Ehrenreich2020} and the system velocity \citep[$v_{sys}=-1.167$ km s$^{-1}$;][]{Ehrenreich2020} we shifted all of the spectra to rest. We then performed a simple 5-$\sigma$ clipping to remove any spurious pixels due to cosmic rays. In order to preserve the relative flux values between different parts of the spectrum while removing any broadband noise or slight deviations in the shape of the blaze functions across observations, we placed all the spectra on a ``common" blaze in a method similar to \citet{Merritt2020}. Finally, we interpolated all the spectra onto a uniform wavelength grid. 

\subsection{Cross Correlation Analysis} 
\label{s:cc}

\begin{figure}
\begin{center}
\includegraphics[width=\linewidth]{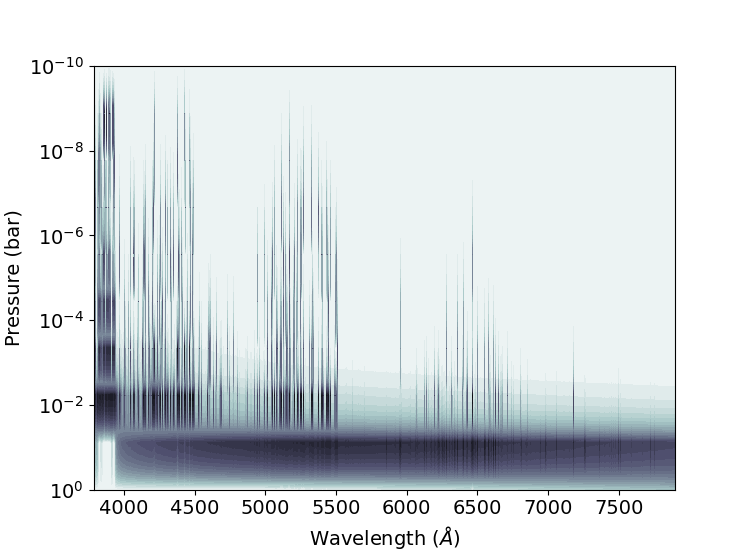}
\includegraphics[width=\linewidth]{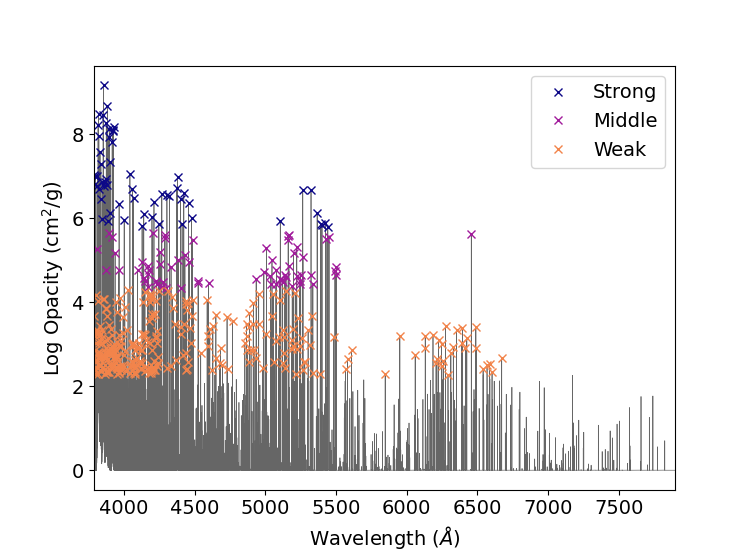}
\caption{\small 
\textbf{Top:} Transmission contribution function, showing the altitude where different Fe lines primarily absorb in the atmosphere of WASP-76 b. Darker colors indicate higher importance of the specific wavelength and altitude in creating the final transmission spectrum, and demonstrates that the Fe line cores primarily absorb at pressures between $10^{-4}$ and $10^{-9}$ bars. The contribution function is calculated using the retrieved exoplanet parameters from \citet{Gandhi2023} and the atmospheric modeling code petitRADTRANS \citep[see ][for details]{Molliere2019}. \textbf{Bottom:} Kurucz line list, showing the opacity of Fe I as a function of wavelength, at a pressure of 1 millibar and assuming a temperature of 2500 K. By comparing the two plots, one can see that the most opaque lines absorb high in the atmosphere, while the less opaque lines absorb lower in the atmosphere as photons will not reach the $\tau$=1 surface until lower depths. The bottom panel also marks which Fe lines are included in each opacity bin. The bin with the most opaque (strong) Fe lines are shown as blue crosses, the middle Fe lines are marked with purple crosses, and the Fe lines that are part of the weak altitude bin are marked with orange crosses.  }
\label{f:MaskBins}
\end{center}
\end{figure}

At this point the cleaned and uniform spectral time series are usually cross correlated with a model atmosphere created using the planet's parameters and one or more molecular or atomic species. This process has proven extremely successful at detecting a range of atoms and molecules in exoplanet atmospheres \citep[e.g.,][]{Snellen2010, Birkby2013, Brogi2017, Hoeijmakers2019}. However, as discussed in detail in \citet{Allart2020}, information on basic absorption line properties such as line amplitude, full width at half maximum (FWHM), or continuum dispersion are lost. Alternatively, a binary mask \citep[e.g.,][]{Allart2020, Pino2018, Pino2020} can be used, which simply consists of a list of absorption line positions for the atom or molecule of interest. The main choice that affects the binary mask is the number of lines used, as some species have hundreds or even thousands of absorption lines. By including a large number of weak absorption lines that are unlikely to penetrate above the continuum of the planet, the signal will be diluted. This mask is then applied to the spectrum and any pixels that fall within the mask are averaged together (the weight of every mask pixel is treated the same and given a value of one in our analysis). In this way, the FWHM or amplitude of the resulting cross correlation function (CCF) from the binary mask is the average FWHM or amplitude of however many absorption lines were included in the mask. Hence, the amplitude of the co-added 1D CCF in the planet's rest frame can be expressed as an average excess absorption from a number of lines in parts per million (ppm) or percent, and is similar to what is reported for single lines resolved by high-resolution spectroscopy. 

Using this binary mask CCF approach, we created multiple masks from the opacity function of the atomic species Fe I. By comparing behaviors between different binary mask CCFs of the same chemical species, we attempt to isolate effects that are due to altitude from effects that may be present when CCFs of different species are compared to each other (i.e., chemical effects). The opacities were downloaded from the DACE opacity database\footnote{\url{https://dace.unige.ch/opacityDatabase}} and created using the open source HELIOS-K opacity calculator \citep{Grimm2021} and the Kurucz atomic line lists \citep{Kurucz2017, Kurucz2018}. We created separate masks for the most opaque lines, the moderately opaque lines, and the weak lines. The most opaque lines become optically thick high up in the atmosphere and they trace the processes occurring at high altitudes, while the resulting CCF from the mask created from the lower opacity lines trace deeper in the atmosphere. This method allows us to directly probe different atmospheric depths which we compare to 3D atmospheric models. 

The opacity model of Fe I, with the absorption lines separated into their three bins is shown in Figure \ref{f:MaskBins}. The lower boundary of the strong lines bin was chosen to approximately map to a pressure of $10^{-5}$ bars, which corresponds to the top of the model atmosphere. As detailed in section \ref{s:gcm}, extending the model to higher altitudes is computationally difficult, and so this strong line bin does not have a model counterpart. For the other two bins, we chose the number of lines for each which led to an even split in signal in the resulting CCFs between the final ESPRESSO spectra and the mask (see SNR values in Table \ref{tab:FeRVs}). To determine the lower bound of the weak lines bin, we looked for the point where adding more Fe lines into our CCF mask no longer increased the resulting signal to noise ratio (SNR), calculated by averaging the CCFs in planet's rest frame and comparing the peak value to noise far from the expected peak. By comparing the bottom and top panel of Figure \ref{f:MaskBins}, the lower bound of the weak line bin approximately maps to a pressure of 1 millibar, which is consistent with the cloud deck location from the retrievals in \citet{Gandhi2022}. While the binary mask CCF approach is less sensitive to parameter choices such as temperature than a model CCF approach due to line shapes not being important, opacity is still temperature sensitive. We find that we need to adjust the values that are used for the bin cutoffs (e.g., bottom panel of Figure \ref{f:MaskBins}) when using opacities of different temperatures, but when the bins are chosen so that the heights of the resulting CCFs are consistent, the derived parameters (radial velocity, FWHM, etc.) are also consistent to within the quoted uncertainties. 

%opacity is temperature sensitive, we find that \textbf{the opaci within the known temperature uncertainty from retrievals \citep[e.g., ][]{Gandhi2022} the lines in each bin are stable to $<5\%$, and therefore do not change our results. 

We cross correlated each night of cleaned spectra with the three different binary masks. While the velocity bins for both nights were the same, the phase coverage and spacing on each night was different. We therefore interpolated the CCF grids onto a uniform phase-space grid, and then co-added the two together, weighting the contribution of each by their SNRs and the number of spectra taken during transit.

\section{GCM modeling}
\label{s:gcm}

\begin{figure*}
\begin{center}
\includegraphics[width=6in]{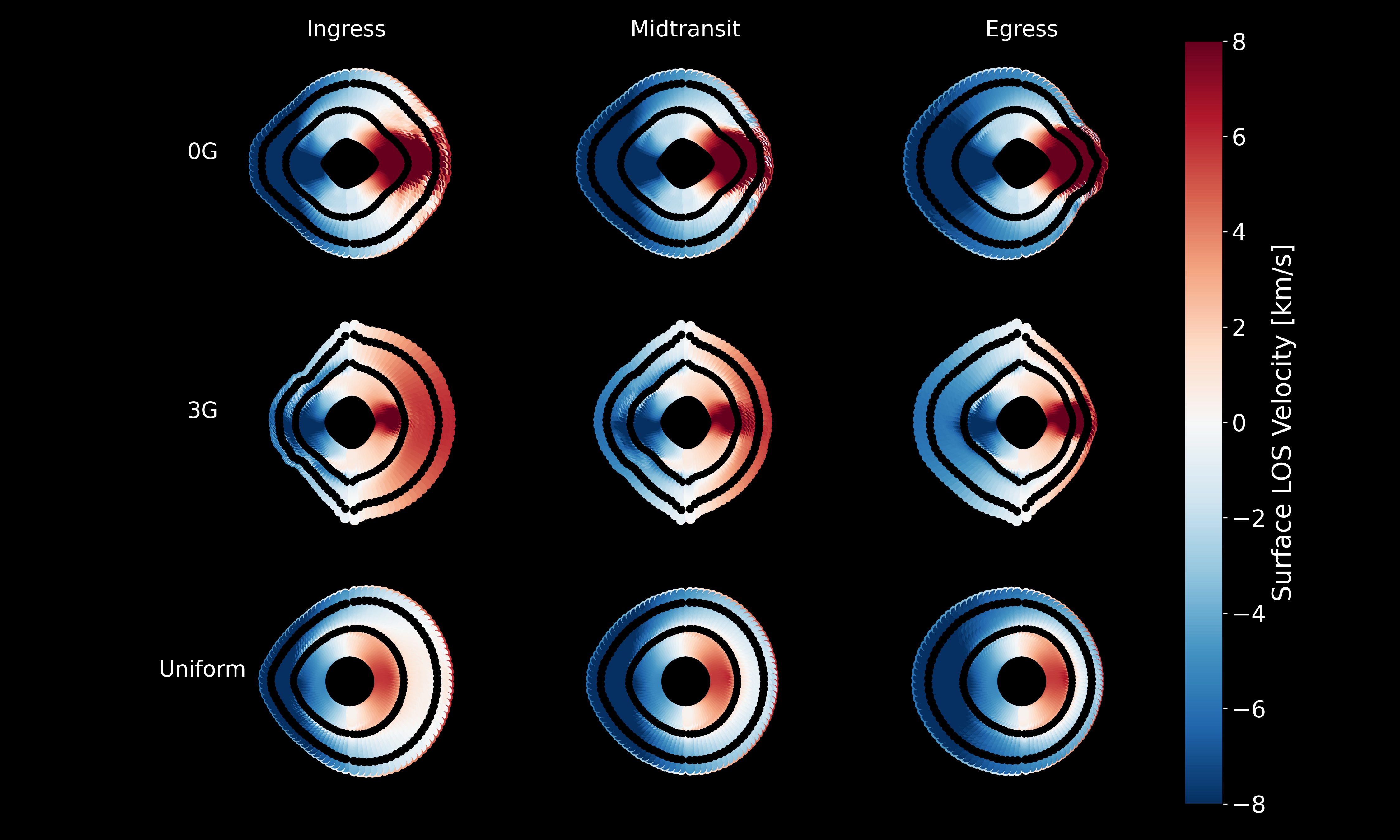}
\caption{\small Line-of-sight (LOS) velocities including the effects of both winds and rotation during ingress, mid transit, and egress for the east and west limbs from the three different GCM models analyzed in this work. Contours of constant pressure (shown in black) correspond to the pressure levels for the medium and weak lines. The inner boundary corresponds to 1 bar and the inner black circle is not to scale. In this projection, the left half of each plot corresponds to the eastern terminator at each phase while the right half corresponds to the western terminator. The limbs are plotted in altitude space and are to scale, with the difference in atmospheric extent between the limbs due to changing scale heights resulting from different temperature structures.However, we note that these models did not include hydrogen dissociation, which can also alter predicted scale heights \citep{TanKomacek2019}. The 3G model has the largest day-night temperature contrast and therefore shows the starkest limb extent contrast.  The two classes of lines are probing different LOS velocities, which are dependent on the choice of magnetic prescription.}
\label{f:GCMMlos}
\end{center}
\end{figure*}

In this work, we post-processed three models of WASP-76b \citep[first published in][where more details and specific numerical parameters can be found]{Beltz2022a}. These are double-grey models using the RM-GCM \citep{Rauscher2012GCM,newradRomanRausher}, with 65 vertical layers evenly spaced in log pressure, from 100 to $10^{-5}$ bars, and a horizontal spectral resolution of T31, corresponding to roughly 4 degree spacing at the equator. In this paper, we focus on the effect of drag and differing drag prescriptions since the main observables we are comparing to are radial velocities and line shapes, which are measurements or winds and are dominated by the treatment of drag. The three models presented differ in their treatment of magnetic effects. The simplest case, the 0G/drag-free model, represents the hydrodynamics only case. The uniform drag model applies a single global drag timescale of $10^4$ s, chosen to match the strong drag case from \cite{TanKomacek2019}. 
Our final model, the 3~G case,  applies an active drag prescription, also known as a ``kinematic MHD" method, and was first applied to hot Jupiters in \cite{RauscherMenou2013}. This method applies a drag on the winds in the east-west direction \citep[as geometrically appropriate for a dipole global field, see][]{Perna2010magdrag} and with a timescale calculated based on local conditions, using the following expression from \citet{Perna2010magdrag}:
\begin{equation} \label{tdrag}
    \tau_{mag}(B,\rho,T, \phi) = \frac{4 \pi \rho \ \eta (\rho, T)}{B^{2} |sin(\phi) | }
\end{equation}
where $B$ is the chosen global magnetic field strength (in this case 3~G), $\phi$ is the latitude, $\rho$ is the density, and $\eta$ the magnetic resistivity, which is a strong function of temperature. The 3~G model is presented here as it most closely matched the \textit{Spitzer} phase curve of the planet \citep{May2021}. Compared to the drag-free models, previous work has shown that the consideration of active drag results in a variety of behaviors not seen in drag free or uniform drag models. The differences most relevant to this paper are as follows:
\begin{itemize}
    \item The active drag model produces a different dayside upper atmosphere flow pattern. The 0~G model and uniform drag model show day to night flow centered on the dayside, with the uniform drag model showing slower wind speeds. Our active model shows flow moving mostly in the north-south direction, up and over the poles \citep{Beltz2022a}.
    \item Increasing the strength of our magnetic field results in a decrease of the hotspot offset (until it rests at the substellar point) and increase the day-night temperature contrast \citep{Beltz2022a}.
    \item Active drag models will show different Doppler shift trends in high-resolution emission \citep{Beltz2022b} and transmission \citep{Beltz2023} spectra, compared to 0~G and uniform models of the same planet.   
\end{itemize}

In Figure \ref{f:GCMMlos} we show line of sight (LOS) velocities including the effects of both winds and rotation for the east and west limbs during ingress, midtransit, and egress. These are plotted as a function of altitude and thus the atmospheric extent is highly influenced by the temperature structure. The two black contours indicate the pressure levels probed by the middle (outer contour, roughly 10 $\mu$bar) and weak lines (inner contour, roughly 1 mbar), respectively. The strongest lines probe pressures not captured by the GCM and are not shown. These differences in velocity structure---between types of drag as well as pressure levels---influence the net Doppler shifts found in the postprocessed spectra (see Section \ref{s:comp_gcm}).  Although there are also varying temperature structures between the models, the winds will dominate the determination of the net Doppler shift.

We then post-processed the two GCMs to generate high resolution transmission spectra at R$\sim$ 450,000. Using ray-striking radiative transfer \citep[described in detail in][]{Kempton2012,savel2022}, we take into account 3D effects such as temperature structure and Doppler shifts from winds and rotations. To generate Fe opacity tables, we assumed solar-abundance equilibrium models \citep{Lodders2003,Fastchem2018} with line lists from \cite{kurucz1995kurucz}.

As the data are cleaned and processed in many steps, which masked or removed some fraction of pixels, we injected the model transmission spectra into the data at the same K$_p$ and a systemic velocity of 100 km s$^{-1}$ (leading to an offset of 101 km s$^{-1}$ from the true planet's signal) in order to perform the same cleaning steps on the models and therefore accurately compare the models to the data. The model transmission spectra were generated for five phases: -0.04, -0.02, 0.0, 0.02, 0.04. For each model transmission spectrum, we interpolated the model onto the same wavelength grid as the data, and then follow convention and simply multiply the data by the interpolated model transmission spectrum for the corresponding phase. We then performed the exact same cleaning steps on the data that have been inject with the model as we did for the data that was not injected with the model \citep[e.g., ][]{Brogi2019, Kesseli2020}.

\section{Results}

We begin by discussing the trends we observe in Fe I in the atmosphere of WASP-76b, as Fe I shows the strongest signal. We then move on to comparing the Fe I trends to trends we detect in other atoms. Finally, we compare the trends we see in the data, with the trends we find from the injected GCM models. 

\subsection{Observed Fe I Trends with Altitude}
\label{s:results_fe}

\begin{figure}
\begin{center}
\includegraphics[width=\linewidth]{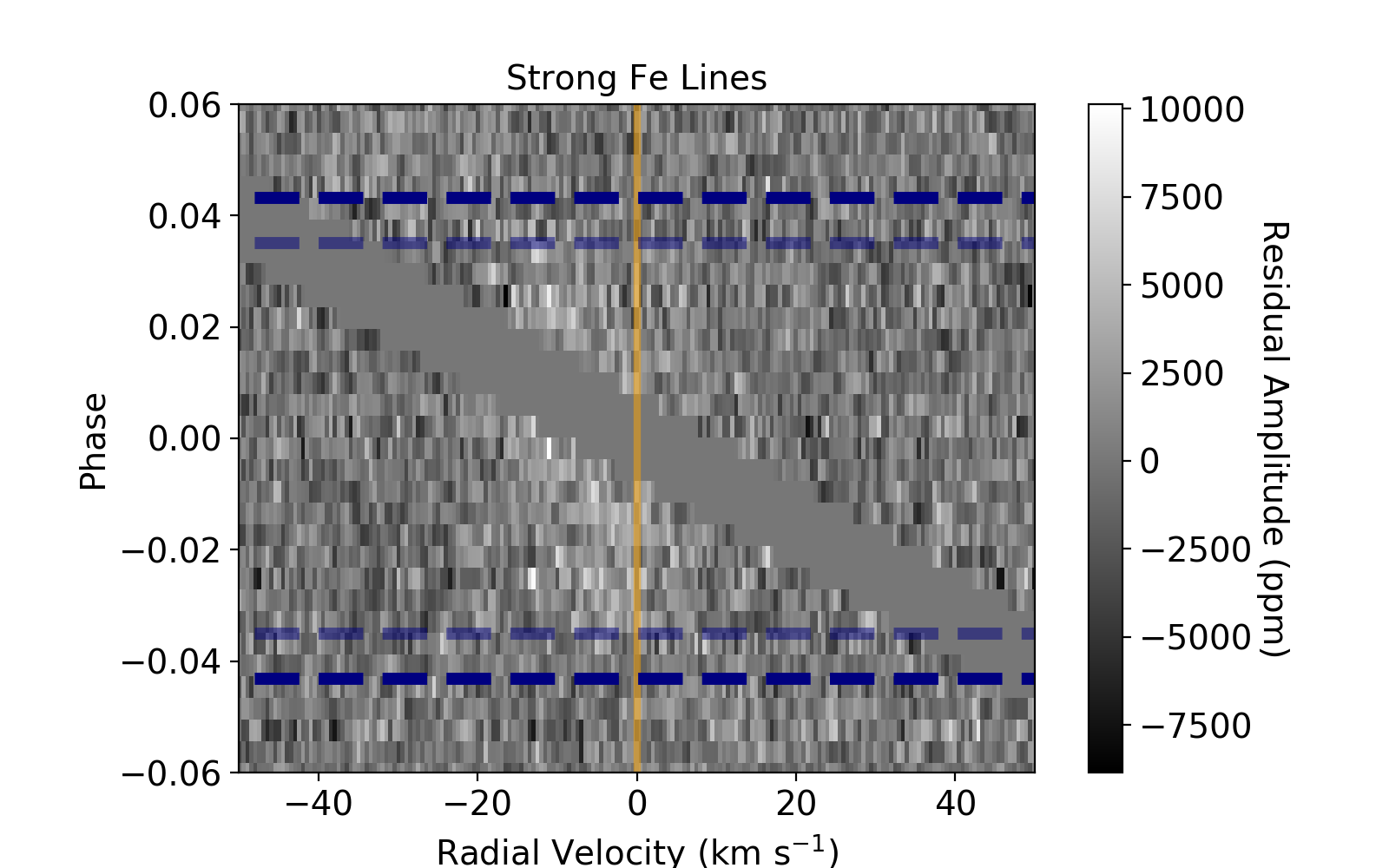}
\includegraphics[width=\linewidth]{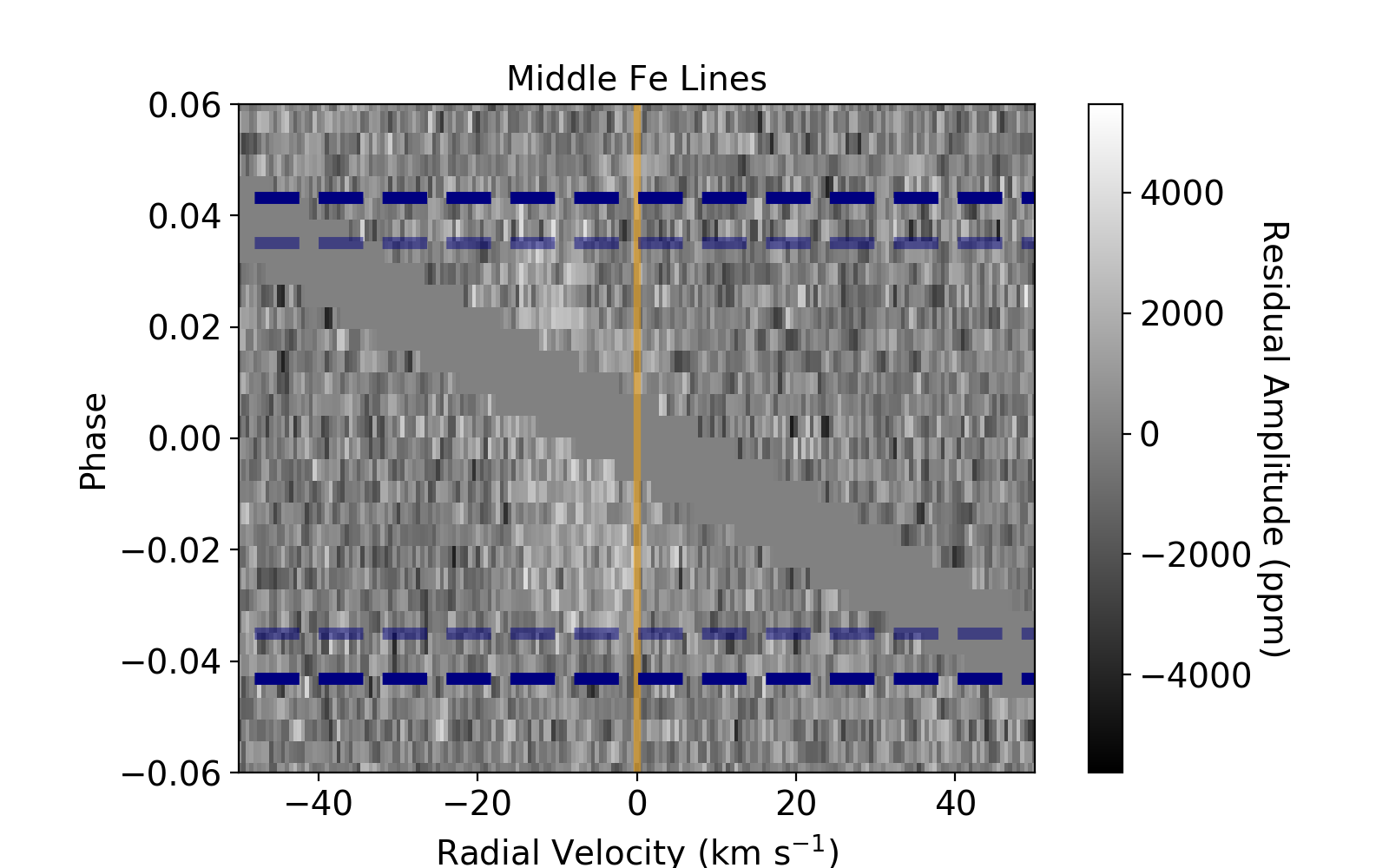}
\includegraphics[width=\linewidth]{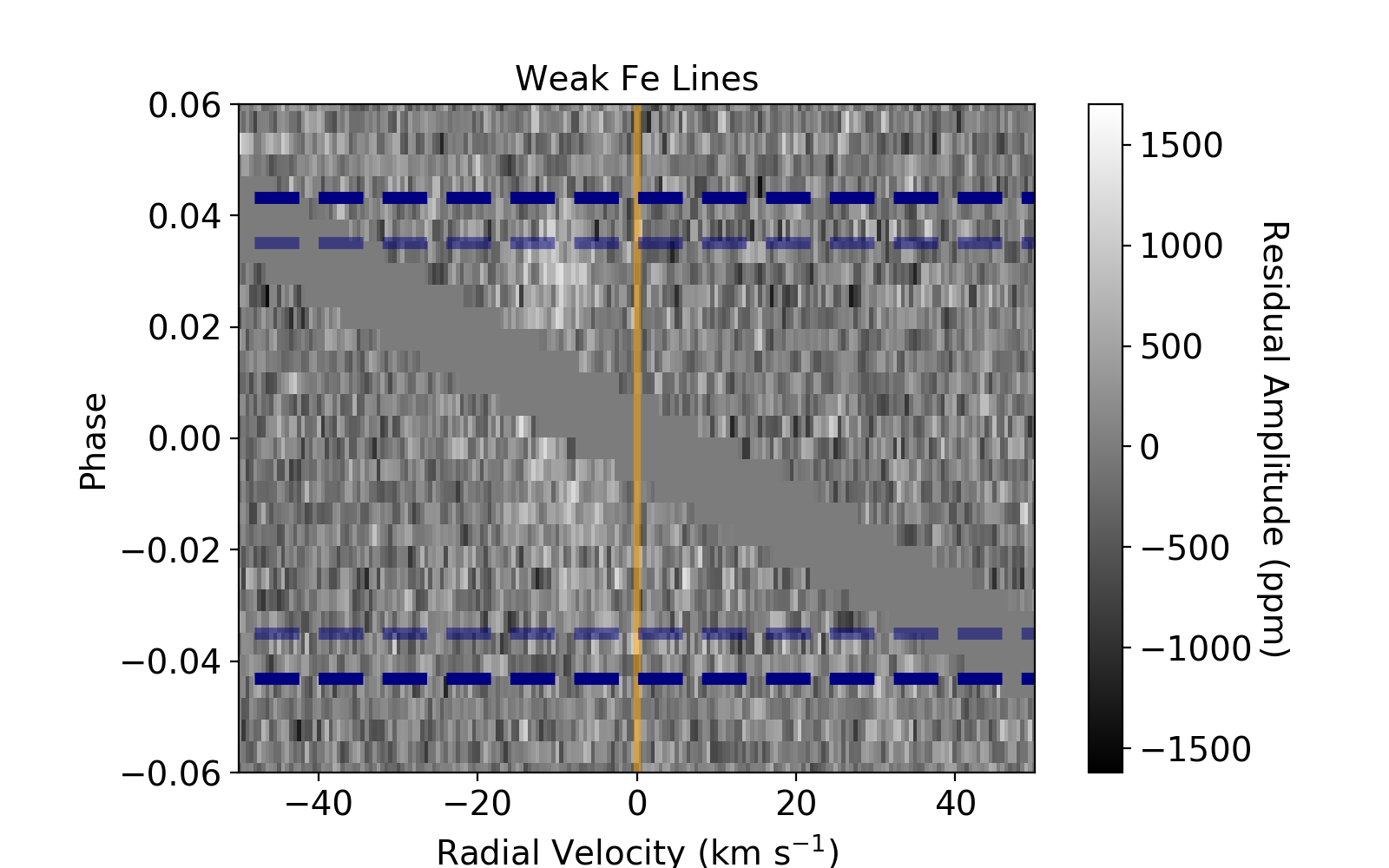}
\caption{\small Binary mask CCF grids for the three different opacity bins from the combined two ESPRESSO datasets. The horizontal lines show the point of first contact/last during transit (opaque navy) and the point of full transit (transparent navy). White residuals indicate excess absorption at the line positions included in each separate binary mask. The grids have been shifted into the planet's rest frame using the well-constrained planet velocity, and so without any atmospheric dynamics we would expect the signal from the planet to lie along the orange line at 0 km s$^{-1}$. We have masked out a region around the host star's rest velocity (diagonal gray stripe) that shows contamination from the Rossiter-McLaughlin effect. In each case we detect a clear signal from the planet with SNRs of 8.73, 7.04, and 5.64 for the strong, middle, and weak bins, respectively.  }
\label{f:2DCCFs}
\end{center}
\end{figure}

\begin{figure}
\begin{center}
\includegraphics[width=\linewidth]{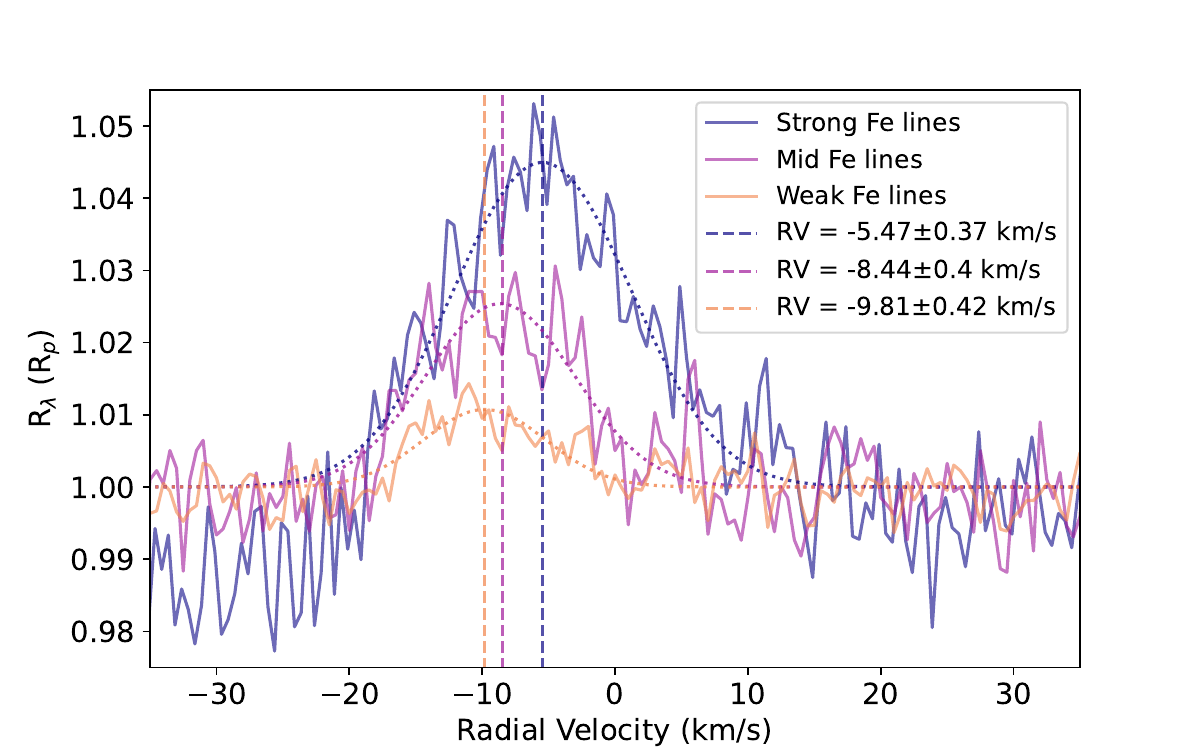}
\caption{\small The same binary mask CCFs for the three different opacity bins (strong, middle, and weak Fe I lines) shown in Figure \ref{f:2DCCFs}, now co-added across the full transit. We show our Gaussian fits to the CCFs with dotted lines and the central radial velocity of the Gaussian for each bin with dashed lines. We find that at each altitude the Fe I signal is blueshifted with respect to the planet's rest frame and also see a significant trend where at higher altitudes (stronger lines) the Fe I signal is less blueshifted. }
\label{f:1DCCF_3bins}
\end{center}
\end{figure}

We created three separate CCF grids from the separate strong-line, middle-line, and weak-line binary masks applied to the combined two nights of ESPRESSO data. Due to the slow rotation of the host star (v $\sin i=1.48$ km s$^{-1}$; \citealt{Ehrenreich2020}) and misaligned orbit of the planet ($\lambda=61.28 ^{\circ}$; \citealt{Ehrenreich2020}), contamination from Doppler shadow due to the Rossiter-McLaughlin effect is confined to a small region around $\pm$10 km s$^{-1}$ in the rest frame of the star. Due to known difficulties in correcting for the Rossiter-McLaughlin effect \citep[e.g.,][]{Casasayas2021b}, we simply choose to mask out this region of the CCF grids. We used the planet's well-known velocity semi-amplitude (K$_p$) of 196.5$\pm1$ km s$^{-1}$ from \citet{Ehrenreich2020}, and shifted all of the cross correlation functions to rest. These results are shown in Figure \ref{f:2DCCFs}, and the presence of Fe I lines in the planet's atmosphere in all three opacity bins is clearly detected as white residuals in the planet's rest frame (along 0 km s$^{-1}$). As expected, the CCF grid created using the strong-line binary mask shows the largest residual amplitudes, meaning that these lines probe higher in the atmosphere. The middle and weak-lined masks show decreasing values in the residual amplitude as they probe deeper in the atmosphere. 

The residual amplitudes in Figure \ref{f:2DCCFs} are related to the planetary radius or number of scale heights above the continuum where the absorption occurs \citep[][]{Tabernero2021, Casasayas2022}. We converted the residual amplitudes to planetary radii using the following equation: 

\begin{equation} 
R_\lambda = \sqrt{1 + h/\delta} R_p , 
\end{equation}

\noindent where $R_\lambda$ is the effective planetary radius above the continuum where the average absorption occurs, $h$ is the measured residual amplitude, and $\delta$ is the white light transit depth of the planet. We also convert this radius to scale heights byassuming a scale height of 1500 km, as reported for the dayside in \citet{Ehrenreich2020}.

We combined each different CCF grid in time by co-adding all of the phases within transit to get three 1D CCFs in the planet's rest frame (Figure \ref{f:1DCCF_3bins}). Consistent with Figure \ref{f:2DCCFs}, the CCF created from the mask that used the weak Fe lines only extends to about 1.01 $R_p$, whereas the strong line CCF extends to 1.06 $R_p$. 
In the absence of any atmospheric dynamics and planetary rotation, we would expect the full phase-combined 1D CCFs to peak in the planet's rest frame, close to 0 km s$^{-1}$, but Figure \ref{f:1DCCF_3bins} shows that at all altitudes the planet's absorption is significantly blueshifted. As the signals are noisy and we do not know the underlying functional form of the CCFs, we simply fit Gaussians to each 1D CCF in the region around the observed signal ($\pm80$ km s$^-1$) using the python curve fitting code, \texttt{lmfit}, to determine the average radial velocity shift and an uncertainty on that radial velocity for each bin (see Appendix \ref{A:unc} for more information on the reliability of these values and uncertainties). We also calculated SNRs by taking the peak value of the Gaussian fit and dividing by the standard deviation far from the peak (outside of $\pm80$ km s$^-1$). The measured radial velocities for each bin are plotted in Figure \ref{f:1DCCF_3bins} and the SNR of each CCF is reported in the caption of Figure \ref{f:2DCCFs}. These measured radial velocities for each CCF show a significant trend where the Fe lines in the lower atmosphere appear to be more blueshifted, and as we look higher in the atmosphere the lines become less blueshifted.

\begin{table}[h!]
\vspace{14pt}
    \centering
    \begin{tabular}{|c|c|c|c|}
    \hline
         & CCF Mask & $\phi_-$ & $\phi_+$\\
      \hline 
      \hline
        SNR & Full & \multicolumn{2}{c|}{13.2}  \\
        \hline
        SNR & Strong & 5.07 & 4.68\\
          & Middle & 5.55 &  5.35\\
         & Weak & 2.84 &  5.62 \\
         \hline
        Height  & Strong & $1.046\pm0.005$ & $1.061\pm0.006$\\
        (R$_p$) & Middle & $1.024\pm0.003$ & $1.039\pm0.004$ \\
         & Weak &  $1.011\pm0.002$ & $1.024\pm0.002$ \\
         \hline
       RV shift  &  Strong & $-3.59\pm0.77$ & $-9.98\pm0.48$\\
       (km/s) &  Middle & $-4.99\pm1.02$ & $-10.69\pm0.44$\\
        &  Weak & $-3.55\pm0.85$ & $-11.12\pm0.34$\\
        \hline
        FWHM  & Strong & $15.5\pm1.8$ & $11.0\pm1.3$\\
        (km/s) & Middle & $18.2\pm2.4$ & $8.9\pm1.0$\\
         & Weak & $9.1\pm2.0$ & $7.4\pm0.8$\\
         \hline
         Height Ratio & Strong & \multicolumn{2}{c|}{$0.76\pm0.09$}\\
         ($\phi_-/\phi_+)$& Middle & \multicolumn{2}{c|}{$0.62\pm0.08$}\\
         & Weak & \multicolumn{2}{c|}{$0.44\pm0.08$}\\
         \hline
    \end{tabular}
    \caption{Measured Fe I CCF parameters for the two different phase bins and three different opacity (altitude) bins. $\phi_-$ consists of phases from $-0.04$ to $-0.02$, while $\phi_+$ consists of phases from $+0.02$ to $+0.04$. We also report the measured SNR for each phase bin and over the entire transit when all the lines are used (Full). All uncertainties quoted are measured using \texttt{lmfit} (see Appendix \ref{A:unc}). The Height Ratios and associated uncertainties were calculated using the best-fit Heights for both phase bins in units of residual amplitudes (ppm), before they were converted to units of R$_p$. The associated uncertainties for the two Heights were then simply propagated to obtain an uncertainty for the Height Ratio. }
    \label{tab:FeRVs}
\end{table}

%We see the clear trend of increasing blueshift during the course of transit that was first seen in WASP-76b  \citep{Ehrenreich2020, Kesseli2021} and later WASP-121b \citep{Borsa2021} and GCMs \citep{Wardenier2021, savel2022,Beltz2023} and is a result of day-to-night winds and the hotter limb coming into view. 

\begin{figure*}
\begin{center}
\includegraphics[width=\linewidth]{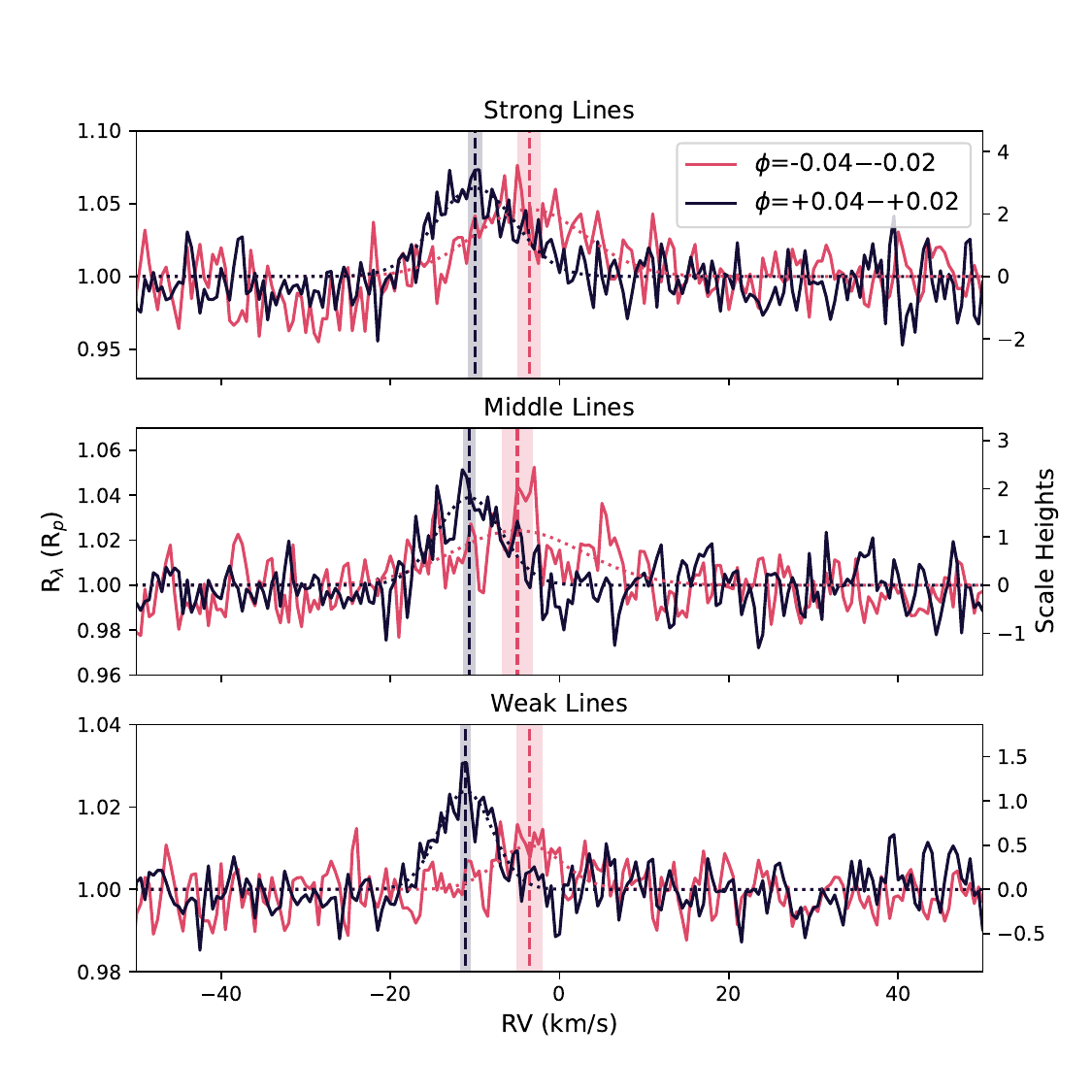}
\caption{\small Same plot as Figure \ref{f:1DCCF_3bins}, but now we have split the each 1D CCF into two temporal bins, phases -$0.04-$-0.02 (pink) and +0.02$-$+0.04 (black), in order to assess the Fe I asymmetry at different altitudes. The dotted lines of the same color show the Gaussians that were fit to each CCF. The vertical dashed lines show the measured peak radial velocities of the Gaussians, and the shaded regions correspond to the 2-$\sigma$ uncertainties in the position of the peak. The radial velocities and uncertainties that are plotted here are all displayed in Table \ref{tab:FeRVs}.}  
\label{f:1DCCF_3bins_split}
\end{center}
\end{figure*}

WASP-76~b shows a strong asymmetry between the absorption depth and the radial velocity shift measured in the first half of the transit versus the second half of the transit \citep{Ehrenreich2020, Kesseli2021}, which can be seen clearly in Figure \ref{f:2DCCFs}. This blueshift evolution over the course of the transit has been explained with GCMs \citep{Wardenier2021, savel2022,Beltz2023} as being due to a combination of day-to-night winds and the hotter east limb (which is more blueshifted from tidally locked rotation) coming into view at the end of the transit. To try to isolate changes that are due to the vertical structure and not the change in viewing angle,  we also split the full transit into two phase ranges that separated the beginning and end of the transit and co-added these two parts of the CCF grid separately (Figure \ref{f:1DCCF_3bins_split}). When we examine Figure \ref{f:1DCCF_3bins_split}, we see that the significant trend of blueshift decreasing with altitude observed in Figure \ref{f:1DCCF_3bins} is no longer clearly observed. The majority of the perceived increasing blueshift is due to the fact that at lower altitudes the signal from the beginning of the transit ($\phi_-$) is drastically diminished and so the combined CCF from the whole transit is influenced more by the $\phi_+$ phases which are more blueshifted. 

We report the measured radial velocity shifts for the three altitude bins, separated by phase, in Table \ref{tab:FeRVs}. We find highly consistent values and trends for the measured radial velocities at the beginning and end of the transit when we compare our values to those from previous studies of \citet{Ehrenreich2020} and \citet{Kesseli2022} even though we use a completely different method to either paper, demonstrating the reliability of the measured velocities and their lack of dependence on reduction method (see also Appendix \ref{A:unc} for more discussion). Looking at how the measured radial velocities change with altitude, we find that the radial velocities at all altitudes are consistent within their uncertainties for the beginning of the transit ($\phi_-$). For the $\phi_+$ bin, we find that the radial velocities show a trend for stronger wind speeds deeper in the atmosphere (i.e. larger blueshifts are measured for the opacity bin that traces the deepest point in the atmosphere). In order to determine the significance of this trend we performed a linear regression using \texttt{scipy.ord}, which was chosen due to its handling of uncertainties in regressions, and find a low-significance trend with a p-value of 0.02 or 2.3-$\sigma$. Only observing a trend during the end of the transit ($\phi_+$) can be explained given that recent GCM modeling \citep{Wardenier2021, savel2022} and phase-resolved retrieval results for WASP-76~b \citep{Gandhi2022} suggest that during the second half of the transit the evening side completely dominates and the morning side of the atmosphere cannot be resolved, while during the beginning of the transit both the morning and evening side of the atmosphere can be observed. Therefore, the $\phi_+$ CCFs originate from a single hemisphere, making it easier to observe any trends within the data. The three bins trace altitudes of $\sim$1.02 to 1.1 R$_p$, or 1 to 4 scale heights, meaning that at most we measure a 1.15 km s$^{-1}$ change in the wind between these altitudes.

We also report the FWHM of the CCFs and the ratio of the heights between the phase bin for the beginning of the transit and the phase bin for the end of the transit in Table \ref{tab:FeRVs}. These values and their uncertainties were also determined by fitting Gaussians to the CCFs shown in Figure \ref{f:1DCCF_3bins_split}. The CCFs for the phase bin encompassing the beginning of the transit are on average 6 km s$^{-1}$ wider than for the phase bin at the end of the transit. This is expected given the results from the GCM modeling and retrievals mentioned above which conclude that the Fe signal at the beginning of the transit originates from a combination of the morning and evening terminators, while the signal at the end of the transit is dominated by the evening side alone. Again, we only see a trend for the $\phi_+$ phases, and measure an increasing FWHM as we probe higher altitudes. As previously mentioned, the majority of the perceived blueshift evolution in altitude shown in Figure \ref{f:1DCCF_3bins} is caused by the CCF signal in the beginning of the transit becoming weaker for the CCF mask that contained the least opaque lines (weak lines), and is reflected in our reported height ratio in Table \ref{tab:FeRVs}, which decreases at lower altitudes. For the lowest altitude bin we find that during the beginning of the transit the Fe lines become optically thick much deeper in the atmosphere than during the end of the transit when the hotter evening side of the atmosphere is in view. 

Finally, motivated by recent modeling work that discussed how asymmetric absorption would manifest in a $K_p$ vs. $V_{sys}$ diagram \citep[e.g., ][]{Wardenier2021} and observations of different offsets for different species in the same $K_p$ vs. $V_{sys}$ diagram \citep{Kesseli2022, Borsato2023, Brogi2023}, we created separate $K_p$ vs. $V_{sys}$ diagrams for the 3 altitude bins and compare the relative offsets from the planet's known position in Appendix \ref{A:KpVsys}. We do not find any clear trend in $K_p$ or $V_{sys}$ offset for the three bins.

\subsection{Observed Trends in Altitude from other Atoms and Ions}
\label{s:results_other}

\begin{figure}
\begin{center}
\includegraphics[width=\linewidth]{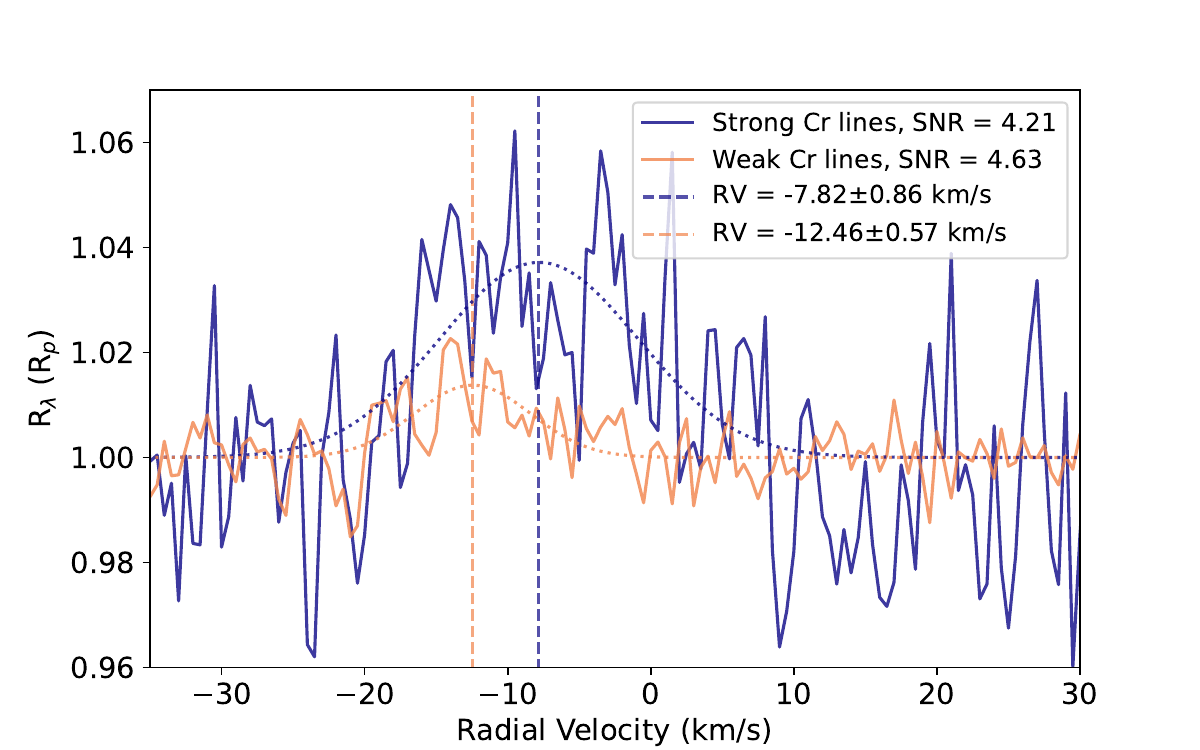}
\includegraphics[width=\linewidth]{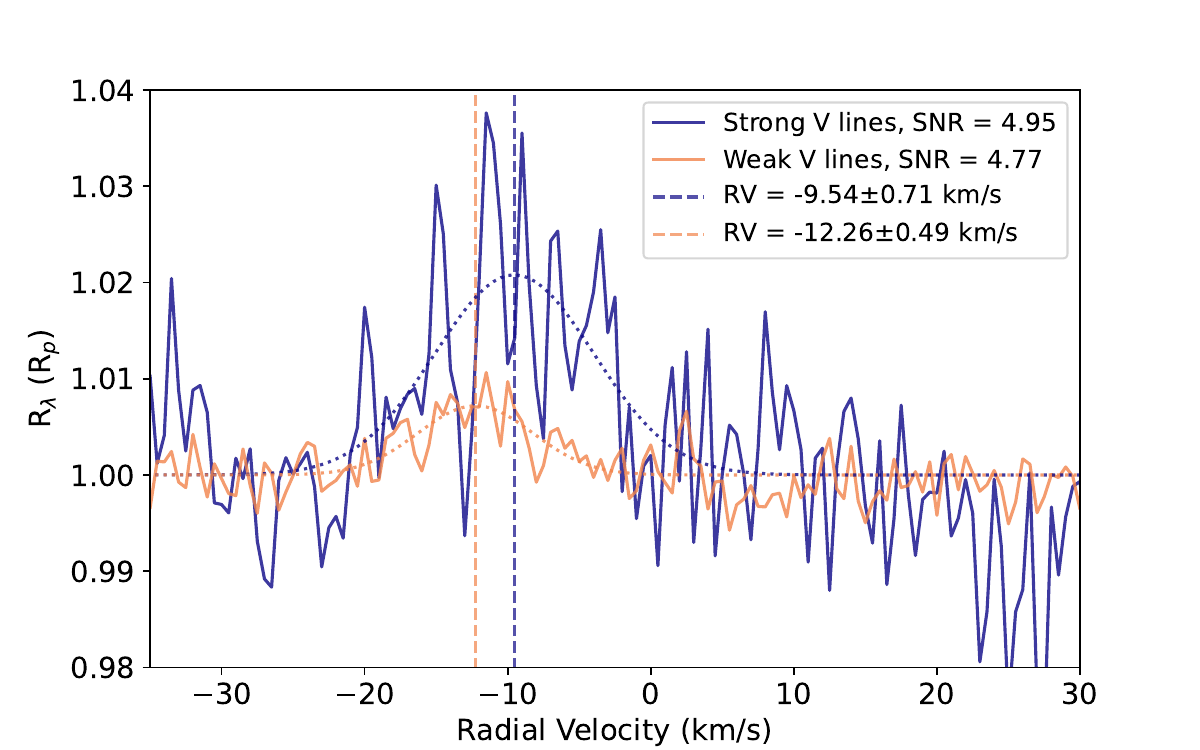}
\includegraphics[width=\linewidth]{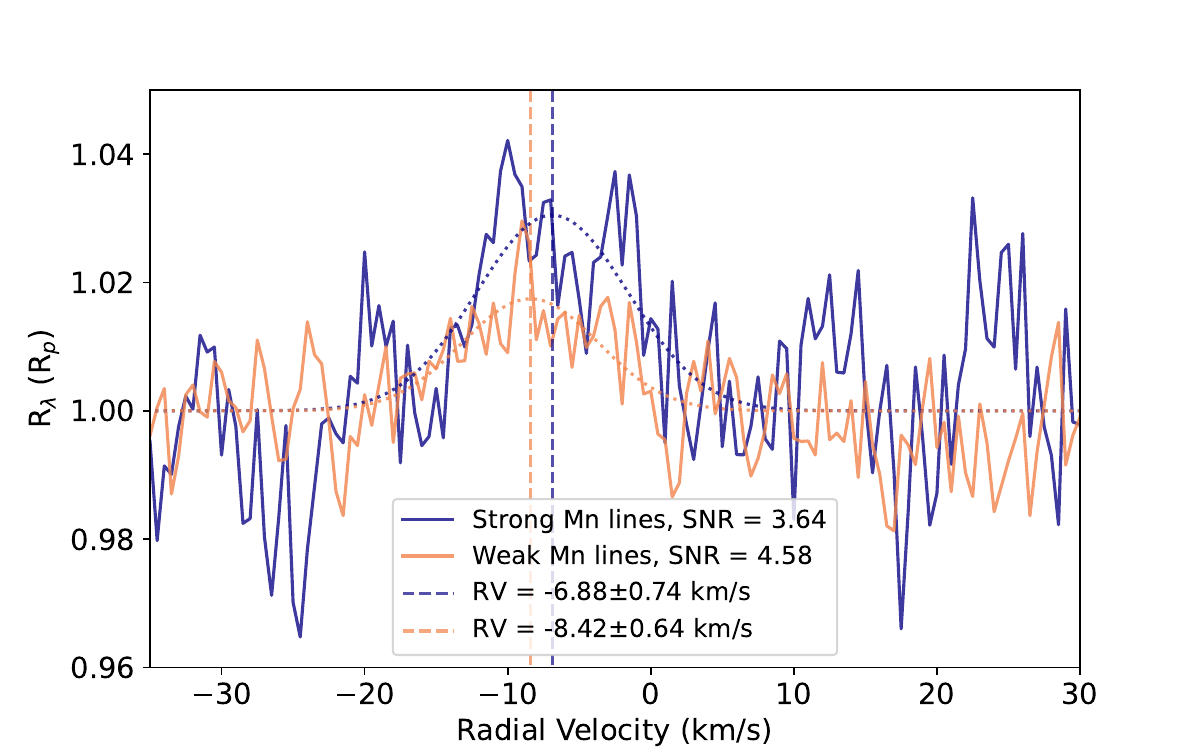}
\caption{\small Binary mask CCFs for Cr I (top), V I (middle) and Mn I (bottom). The CCF masks have been split into two different opacity bins and co-added in the planet's rest frame in a similar manner as Figure \ref{f:1DCCF_3bins}. As in Figure \ref{f:1DCCF_3bins}, we have also plotted the Gaussian fits to each CCF with dotted lines of the same color, and the peak radial velocity of the fitted Gaussian as a vertical dashed line. Even though the CCFs are noisier than the Fe I CCF, we still measure a SNR$>$3.5 in each case, and $>$4.5 in most cases. For all three species we observe the same pattern where the CCFs created using the stronger lines are less blueshifted than the CCFs created using the weaker lines. }
\label{f:OtherSpecies}
\end{center}
\end{figure}

Fe I shows the strongest signal of any of the detected atoms or ions in WASP-76~b \citep{Kesseli2022}, and so we focus our findings on Fe I. However, we also checked whether any of the trends seen in Fe I were also present in the absorption signals from other atoms (Cr I, V I, and Mn I) that showed significant absorption in \citet{Kesseli2022}. For each atom we created two masks (instead of 3 as was done for Fe I) since the signals were not as strong and using more bins would result in less robust detections. One mask contained the most opaque lines while the second mask contained lines that were less opaque. We split the bins at a point where the two masks led to SNRs in the resulting CCFs that were roughly equal. The binary mask CCFs for the two opacity bins and three different atoms, co-added in the planet's rest frame are shown in Figure \ref{f:OtherSpecies}. 

We find the same trends in all three other species that we see in the analogous plot for Fe I (Figure \ref{f:1DCCF_3bins}), and in making the same $K_p$ vs. $V_{sys}$ plots for these species as we did for Fe I, we do not find any clear trends with altitude either (Appendix \ref{A:KpVsys}), reinforcing both conclusions drawn from Fe alone. Figure \ref{f:OtherSpecies} demonstrates that for these three other species the CCF made using the mask of weaker lines shows a larger blueshift than the CCF made using the mask of stronger lines. For Cr I and V I, we measure a significant ($>$3-$\sigma$) difference between the radial velocities of our two altitude bins. Mn I has the lowest SNR of the three trace species, and so even though we find the same blueshifting pattern, the difference between the two measured velocities is less than 2-$\sigma$. The consistency of the observed pattern in radial velocity with altitude for multiple species gives credence to the trend we observe with Fe I. It also demonstrates that the physical mechanism causing the signal to be weaker at lower altitudes during the beginning of the transit (decreasing height ratio trend) is not unique to Fe I, and therefore likely is a process that acts on a global scale (e.g., temperature or wind structure) and not on a single species (e.g., ionization, condensation). 

\subsection{Comparison with GCM models}
\label{s:comp_gcm}

\begin{figure*}
\begin{center}
\includegraphics[width=\linewidth]{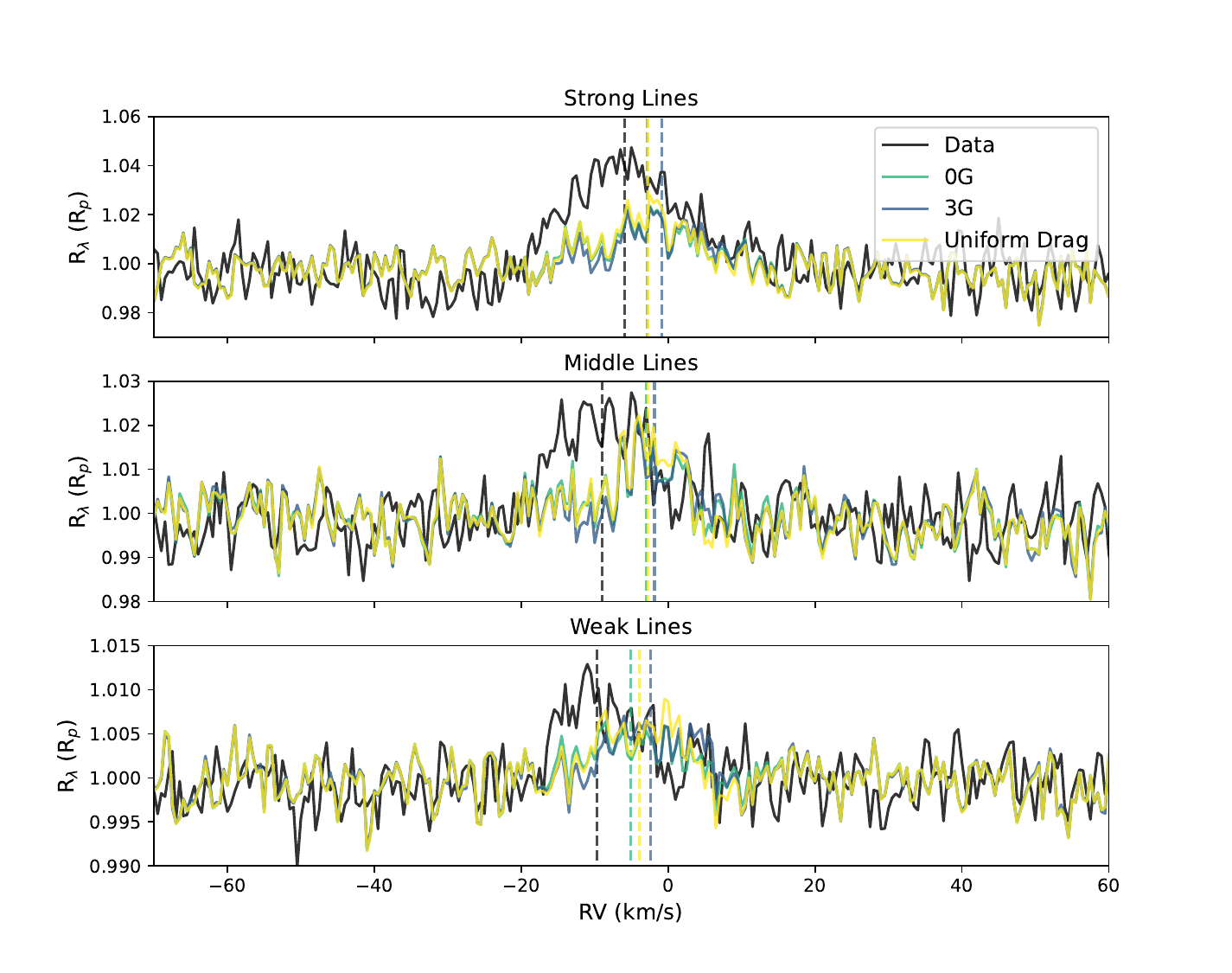}
\caption{\small Comparison between the 1D, phase averaged CCFs for the real atmospheric signal (black solid line) and the three different model atmospheres described in Section \ref{s:gcm} that have been injected into the data (colored solid lines). We also plot the central radial velocity that was obtained by fitting a Gaussian to each CCF with a dashed vertical line. We find that the models underestimate that Fe signal strength at all altitudes, but that this underestimation is worst at the top of the atmosphere, where the Fe in the planet likely is absorbing above the top of the model atmosphere (10$^{-5}$ bars), therefore we do not perform any comparisons between the GCM models and the Strong Line bin for the remainder of the paper to avoid biasing our analysis. We also see that in all cases the models predict the signal to be less blueshifted than the data show. }
\label{f:modelInject}
\end{center}
\end{figure*}

\begin{figure*}
\begin{center}
\includegraphics[width=\linewidth]{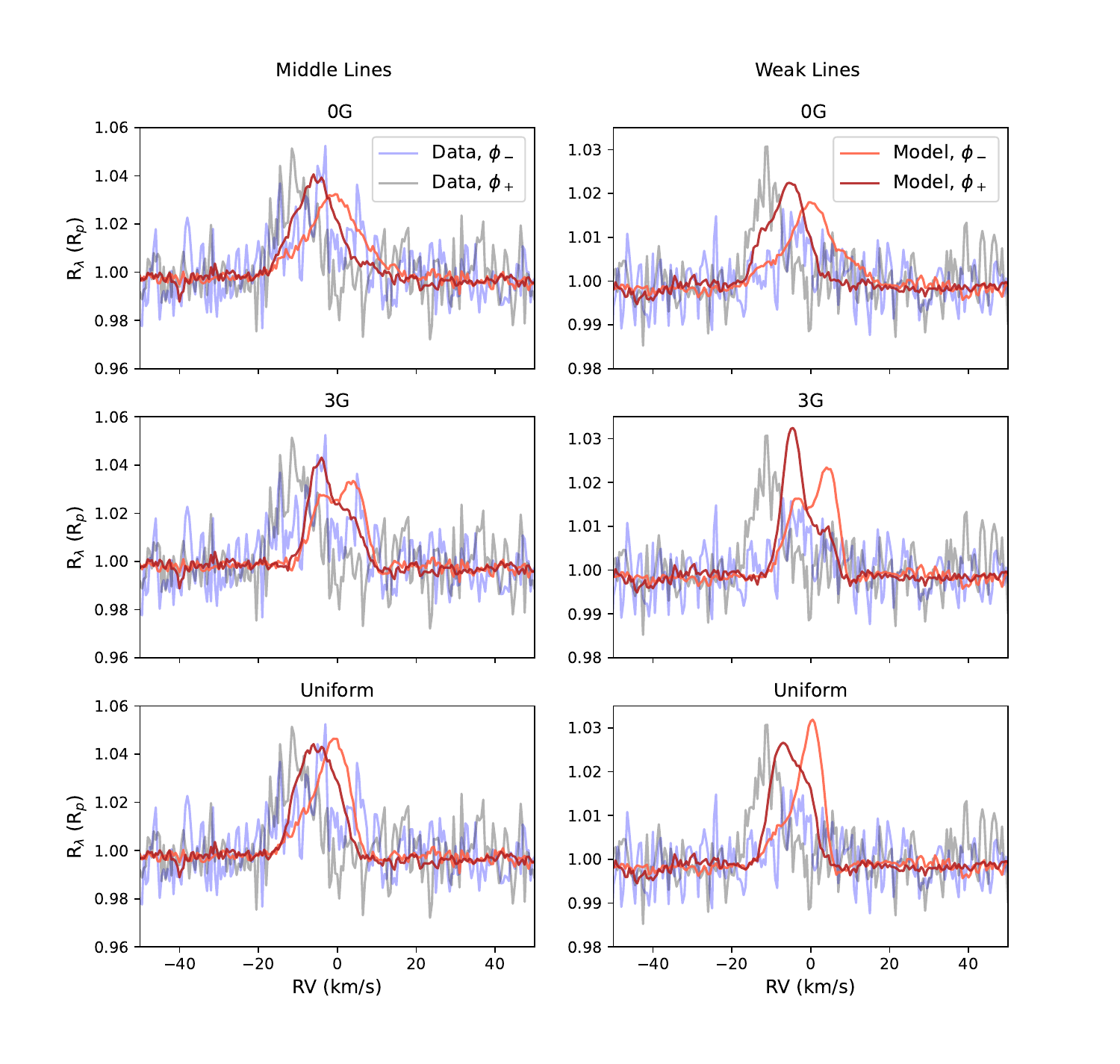}
\caption{\small CCFs where the three different models are injected into the data at many times the expected signal strength to create 'zero noise model CCFs'. As with the data, the CCFs are created with only the middle Fe I lines (left 3 panels) and only the weak Fe I lines (right 3 panels). The CCFs are also split into the phases covering the beginning of the transit ($\phi_-$, light red) and end of the transit ($\phi_+$, maroon). The real Fe I signal from the planet is also plotted in the same manner in transparent blue ($\phi_-$) and black ($\phi_+$). }
\label{f:modelInject_x20}
\end{center}
\end{figure*}

We next aimed to use GCMs to help us interpret these trends and better understand the physical processes occurring in the atmosphere of WASP-76~b. To that end, we compared the measured radial velocities and line shapes of the injected models to the data to determine which drag prescription best fits the data. We injected the different models into the ESPRESSO data at a systemic velocity of +100 km/s so that the true signal and the injected signal would not overlap. We tested injecting the signal at different velocities and found that any differences were within the uncertainties. 

We found that all three models underestimate the strength of the Fe signal at all altitudes (see Figure \ref{f:modelInject}), but that the mismatch with the bin that contains the strong lines was the worst (the models only reach a radius of 1.02 R$_p$ while the data reach 1.05 R$_p$). The top of the atmosphere for the models occurs at 10$^{-5}$ bars, and so this significant underestimation from the models is likely due to the strong Fe lines absorbing at altitudes above 10$^{-5}$ bars in the planet's atmosphere. The top boundary of the GCM was chosen to be 10$^{-5}$ bars, as the the lower atmospheric densities result in numerical instabilities due to extremely short timescales. This upper boundary was also enforced in the post-processing, which is why the CCFs of the GCMs are much weaker for the strong lines. Additionally, non-LTE effects and photoionization --- which are not modeled by the GCM --- play stronger roles in the extended atmosphere \citep{Borsa2021_kelt,Fossati2023,Young2024}. Therefore, we focus the rest of our comparisons on the bins containing the middle and weak Fe lines, as these probe pressure levels covered by our GCMs. Even at lower altitudes, the models still underestimate the signal strength, which may be due to non-solar abundances of Fe -- an assumption made in the post-processing routine -- or differences in temperature structure compared to the true planet.

\begin{figure}
\begin{center}
\includegraphics[width=\linewidth]{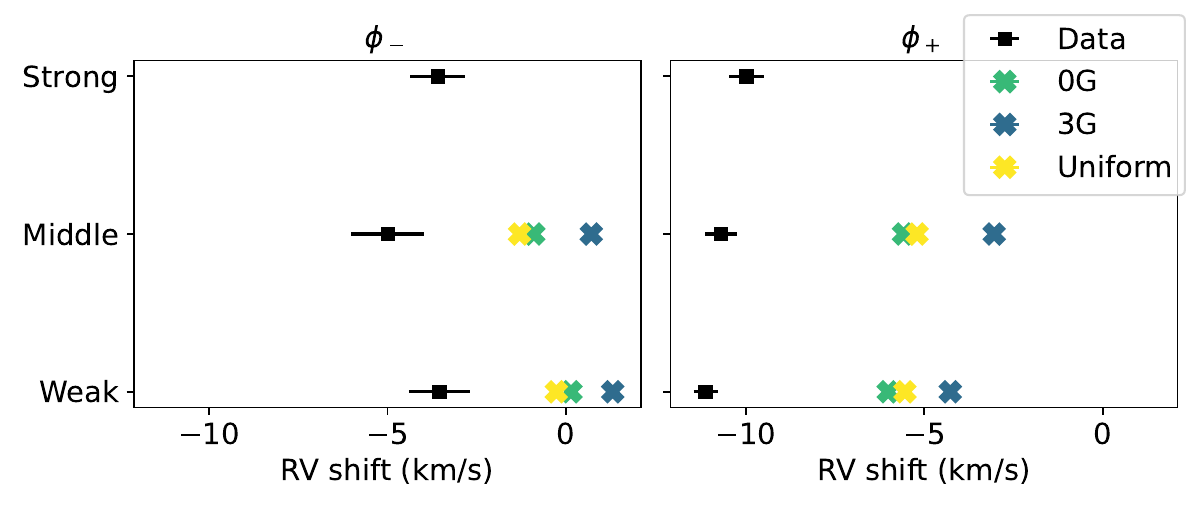}
\includegraphics[width=\linewidth]{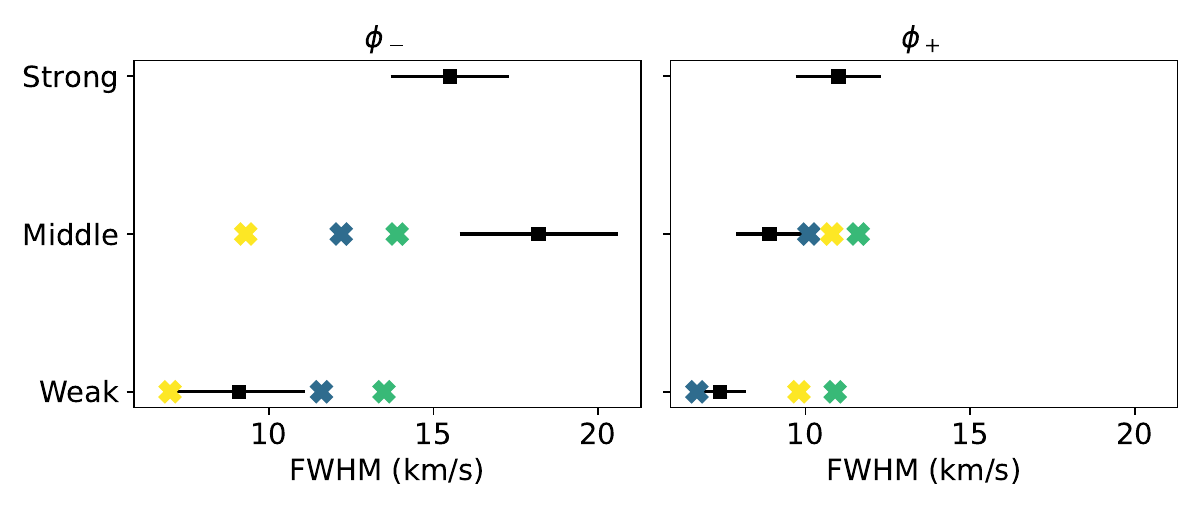}
\includegraphics[width=\linewidth]{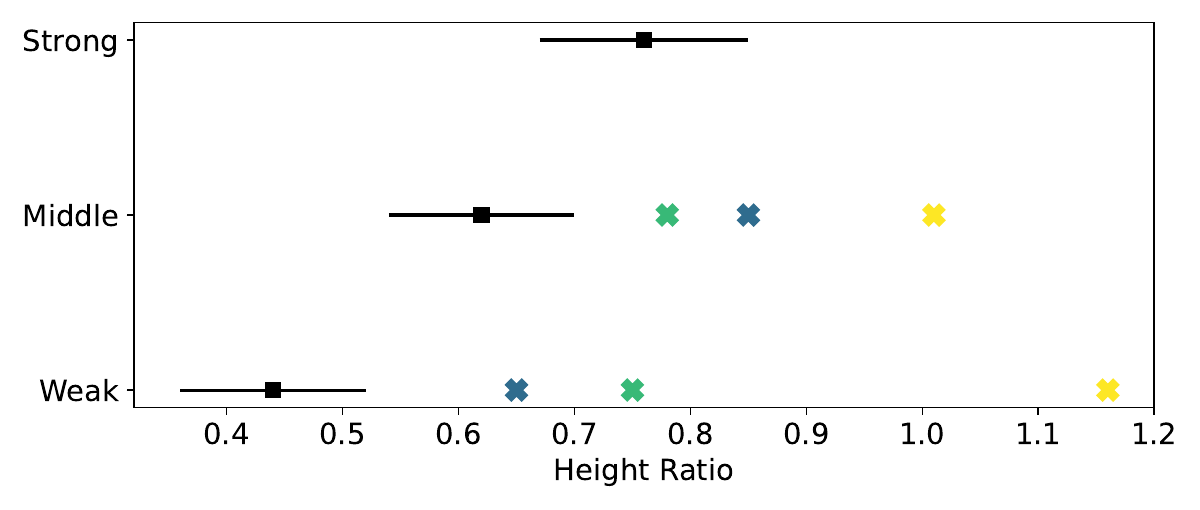}
\caption{\small  Comparison between the CCF parameters for the three different models (0G, 3G, uniform) and the data. The values plotted here can be seen in Table \ref{tab:FeRVs}, for the observed trends in the data, and Table \ref{tab:Models} for the three models. The 1-$\sigma$ uncertainties of both the models and data from their corresponding Tables are plotted in both of the two top panels, but the model uncertainties are smaller than the size of the markers as these uncertainties are measured on the 'zero noise model'. The top panels shows the peak radial velocity shifts for the beginning of the transit ($\phi_-$, left) and end of the transit ($\phi_+$, right) at the three different altitude bins. As the models do not extend to pressures below $10^{-5}$ bars, where the strong Fe line absorb, there are no measurements for the models in the strong line bin in any of the plots. The middle panels are the same, except instead of radial velocity, they show the measured FWHMs of the CCFs. The bottom panel shows the ratio of the height of the CCFs from the beginning of the transit ($\phi_-$) divided by the height at the end of the transit ($\phi_+$). }
\label{f:stats_plot}
\end{center}
\end{figure}

Since our modeled spectra do not create a signal that is as strong as the data, and therefore have even lower SNRs ($\sim2.5-3.5$ for the models and $>5$ for the data), in order to better compare the model to the data we inject the model into the data using standard injection and recovery methods \citep[e.g., ][]{Snellen2010, Kesseli2020, Hoeijmakers2024} at 20 times the expected strength to create `zero noise model' CCFs. We treat these CCFs in the same way as the data and split them into different phase bins, fitting Gaussians to each CCF in order to measure the same CCF parameters for each model as we measured for the data. These newly measured CCF parameters are given in Table \ref{tab:Models}. More complex structure (e.g., double peaks) is now apparent in the CCFs due to the drastically increased SNRs, and demonstrate how this binary mask CCF method preserves line shapes and therefore information about atmospheric wind structures. By comparing Figure \ref{f:GCMMlos}, we can see how differing wind patterns manifest in the CCFs. Despite the more complex structure, we still fit a single Gaussian to the models in order to treat the models and data in a consistent manner and derive parameters that can be compared in a one-to-one manner. The `zero noise' model CCFs are plotted along with the real data CCFs in Figure \ref{f:modelInject_x20}. We also plot all of the summarizing quantities from the two Tables in Figure \ref{f:stats_plot} to enable an easier visual comparison between the models and data. Each model shows unique features that we will attempt to match to the data.

By comparing the CCF parameters for all the models, we see that all of the GCM models reproduce the observed pattern of more blueshifted CCFs in the second half of the transit, which naturally arises from the combination of tidally locked planet rotation and the hotter eastern limb coming into view during the later part of the transit (see Figure  \ref{f:GCMMlos}). As the transit progresses the blueshifted contours become more dominant and are surrounded by regions of higher temperatures. 

Again looking at Table \ref{tab:Models} and Figure \ref{f:stats_plot}, we see that all of the models follow the clear trend in the data that the FWHMs at both phase bins increase with altitude (larger FWHMs for the middle Fe line bin than the weak line bin). We also see that all the models follow the tentative trend that we found in the data where for the $\phi_+$ phases, the degree of blueshifting increases as we probe deeper in the atmosphere (weaker Fe I lines). Examining Figure \ref{f:GCMMlos}, the 3D wind structures at the pressure contours vary significantly, which contribute to the differing Doppler shifts found at these two pressures. This effect is a result of the complex, multidimensional interplay between winds, temperature, atmospheric circulation. The equatorial jet is more pronounced at all longitudes at the deeper pressure level probed by the weak lines resulting in a narrower FWHM and a stronger average blueshift at post transit phases. At the higher level probed by the middle lines the wind structure is far more spatially variable, resulting in a wider FWHM and a smaller average blueshift at post transit phases (even though the winds speeds are typically larger at this altitude).

%For the middle line bin, the difference between the RV shift at the beginning and end of the transit (i.e., RV shift $\phi_-$ $-$ RV shift $\phi_+$) is 5.7 km s$^{-1}$ for the observed data, while the 0G model has the closest match with a difference of 4.68 km s$^{-1}$. The Uniform model and 3G model show RV shift differences of 3.9 and 3.75 km s$^{-1}$, respectively. 

We find that both the 0G and 3G models show significantly larger FWHMs for the first half of the transit than the second half of the transit, as is seen in the data. Conversely, the uniform model shows larger FWHMs for the second half of the transit, which is disfavored by the data. The uniform drag model also shows a stronger signal at the beginning of the transit and a weaker signal at the end of the transit (represented in Table \ref{tab:Models} by a height ratio greater than one). 
Both the 0G and 3G models reproduce the observed trend in the data of a weaker signal at the beginning of the transit. 
The uniform drag model applies the same drag at every spatial point on the planet, which causes less asymmetry than the other two models, and therefore a less dominant evening side of the planet. The signal is therefore weaker and has a wider FWHM due to contribution from both hemispheres. This is evidence that the uniform drag model does not accurately reproduce the observations and that a more complex treatment of drag is necessary. 

\begin{table*}[]
\vspace{14pt}
    \centering
    \begin{tabular}{|c|c|c|c|c|c|c|c|}
    \hline
         & CCF Mask & 0G ($\phi_-$) & 0G ($\phi_+$) & 3G ($\phi_-$) & 3G ($\phi_+$) & Uniform ($\phi_-$) & Uniform ($\phi_+$)\\
      \hline 
      \hline
       RV shift  & Middle & $-0.90\pm0.12$ & $-5.58\pm0.1$ & $0.71\pm0.14$ & $-3.04\pm0.11$ & $-1.3\pm0.08$ & $-5.20\pm0.09$\\
        &  Weak & $0.14\pm0.12$ & $-6.01\pm0.10$ & $1.31\pm0.14$ & $-4.27\pm0.06$ & $-0.27\pm0.06$ & $-5.54\pm0.08$ \\
        \hline
        FWHM  & Middle & $13.9\pm0.2$ & $11.6\pm0.9$ & $12.2\pm0.3$ & $10.1\pm0.3$ & $9.3\pm0.2$ & $10.8\pm0.2$\\
         & Weak & $13.5\pm0.3$ & $10.9\pm0.2$ & $11.6\pm0.3$ & $6.7\pm0.2$ & $7.0\pm0.2$ & $9.8\pm0.2$\\
         \hline
         Height Ratio & Middle & \multicolumn{2}{c|}{0.78} & \multicolumn{2}{c|}{0.85} & \multicolumn{2}{c|}{1.01}\\
         & Weak & \multicolumn{2}{c|}{0.75} & \multicolumn{2}{c|}{0.65} & \multicolumn{2}{c|}{1.16}\\
         \hline
    \end{tabular}
    \caption{Measured Fe I CCF parameters for the three different models that have been injected into the data. As with the data, we split the CCFs into two different phase bins and different opacity (altitude) bins. In the same manner as the data, the values and uncertainties reported here are from fitting Gaussians using \texttt{lmfit}. }
    \label{tab:Models}
\end{table*}

Between the 0G and 3G models, we find some instances where the 0G models seems to be preferred and some where the 3G model seems to be preferred. In the data, we find a decreasing height ratio as deeper pressures are probed, which means that the signal from the $\phi_+$ bin become stronger compared to the $\phi_-$ bin as one looks deeper in the atmosphere. The 3G model shows a decreasing height ratio, like the data, while the 0G model has an almost constant height ratio at the two altitudes. This trend is likely influenced by multiple concurrent physical effects. In our GCMs, the 3G model has the largest day-night temperature contrast, resulting in the largest scale height difference between egress and ingress. The model also shows a jet that is more localized to the lower atmosphere than the 0G case, potentially leading to this large scale height difference being more localized to the lower atmosphere. Other processes, such as clouds or hydrogen dissociation (though not included in these models) could also likely influence this trend, and are discussion more in Section \ref{s:disc}. 
On the other hand, the 0G model matches the measured RV shifts the best, in that it exhibits the strongest blueshift for the $\phi_+$ phases, and the difference between the RV shift between the two phase bins is largest. This is because the 0G model has no additional drag, resulting in a stronger super-rotational jet. It is important to note that no GCM models have been able to reproduce the magnitude of the measured RV shifts found on WASP-76~b, and so there is likely some missing physics in the GCMs. 
Even though all of the models underpredict the observed blueshifts, both the 3~G and 0~G models follow the trend of increasing blueshifts as the transit progresses from $\phi_-$ to $\phi_+$, and increasing blueshifts deeper in the atmosphere.

\section{Discussion}
\label{s:disc}

From our comparisons, we are able to show that applying drag uniformly to the atmosphere is not sufficient to fit the data and we are able to reject the uniform drag model. Previous comparisons to high resolution data found that models with weak drag were required to fit the asymmetric shape of the Fe CCF \citep{Wardenier2021, savel2022}. Conversely, low resolution \textit{Spitzer} phase curves showed little phase offset, and therefore required models with significant drag, therefore setting up a tension between the two methods. As the previous GCM modeling results used a uniform drag prescription \citep[see][]{Wardenier2021, savel2022}, we suggest that the poor fits for the drag models could have been due to the spatially uniform application of drag, and find that the inclusion of \textit{spatially varying drag} is necessary to best fit these high resolution data. From our analysis it is initially not obvious whether the 0~G or the 3~G model is the better fit, as both of the models are able to fit many of the trends. However, some of the first GCM modeling work for hot Jupiters \citep{Kempton2012} predicted that the trend of increasing blueshifts for deeper pressures---which is shown by the data---could be indicative of magnetic effects. Finally, it is notable that the 3~G model was also able to fit the low resolution \textit{Spitzer} phase curves better than the 0~G model \citep{May2021,Beltz2022a} meaning that our \textit{single} kinematic MHD model fits both high resolution and low resolution data, which is a significant feat. As we have not done a full parameter sweep of varying field strengths or magnetic topologies, we are not claiming that the magnetic field strength of this planet is 3G, but take the agreement with both phase curves and high-resolution data as potential indication of magnetic effects shaping the environment.  
%The number of 3G is not claim we are trying to make, but the fact that our drag is spatially varying is the important take away. 

All of the models underpredict the wind speeds, and previous GCMs of this planet have been unable to match the magnitude of the phase-resolved Doppler shifts \citep{Wardenier2021, savel2022,Beltz2023}. \citet{savel2022} suggested this discrepancy between observed radial velocities and those predicted by GCMs could be due to imprecision/uncertainties in the ephemeris and eccentricity, but recent precision measurements of the eccentricity and ephemeris of WASP-76~b from \citet{Demangeon2024} have shown that orbital uncertainties are not to blame for inconsistencies with GCMs. It therefore seems that there is some missing physics that has not been taken into account for the GCMs that is needed to completely fit the WASP-76~b data. As of now it is unclear whether this problem of underpredicting planetary radial velocities is widespread, as there are few planets with this type of measurement, but initial wind measurements in a sample of 6 ultra-hot Jupiters found that WASP-76~b had the highest wind speed \citep{Gandhi2023}.

Future work should explore in more depth why GCMs can underpredict wind speeds in HRS. Running GCMs with lower hyperdissipation and/or higher resolution may lead to higher model wind speeds. In addition, including physical processes such as hydrogen dissociation, clouds, and magnetic drag concurrently should be ran to better physically understand the atmosphere of this planet. Hydrogen dissociation would reduce the day-night temperature contrast \citep{TanKomacek2019} and the corresponding difference in scale height. Clouds, which may be present on one or both terminators, can increase the net blueshift, particularly during ingress \citep{Savel2023}. Understanding how all of these processes act together and influence measured observables will shed light on which are the most important in future exoplanet modeling and the physical mechanisms occurring in planetary atmospheres. 
%We leave it to future studies to look in more depth at what this missing physics is... clouds, dissociation, etc. how these other mechanisms affect wind speed,

\section{Conclusions}
\label{s:conc}

Using two ESPRESSO transits of WASP-76b, we explored how winds and dynamics in the planet's atmosphere changed as a function of vertical altitude, and compared these results to the output of state-of-the-art global circulation models (GCMs). To resolve the vertical structure of the atmosphere, we created binary masks containing Fe I lines in three different opacity bins. The bin with the Fe I lines that were the most opaque probes the highest layers of the atmosphere. The bins with less opaque Fe I lines probe lower in the atmosphere since light can travel through deeper layers of the atmosphere before it becomes optically thick. We then cross correlated these binary masks with the data and explored trends in the cross correlation functions. By cross correlating with a binary mask we are able to preserve information on the true line shapes, and therefore accurately extract the average height in the atmosphere where the signal arises and full width at half maximum (FWHM) of this signal. As the Fe I signal is famously also known to show an asymmetric velocity shift from the start of the transit to the end of the transit \citep{Ehrenreich2020, Kesseli2021, Kesseli2022}, we also split up the cross correlation functions (CCFs) into two phase bins. 

We find that at all heights, the phase bin for the second half of the transit ($\phi_+$) is more blueshifted than the first half of the transit ($\phi_-$), and that the asymmetry first found by \citet{Ehrenreich2020} and confirmed in \citet{Kesseli2021} persists at all altitudes. We also see that at all altitudes the FWHM for the phase bin encompassing the beginning of the transit is significantly larger than the phase bin for the end of the transit. We take this as evidence that at the end of the transit the signal is dominated by only one side of the planet (the hotter evening side), whereas at the beginning of the transit the signal is a combination of the beginning and end of the transit, consistent with phase-resolved retrievals of WASP-76b \citep{Gandhi2022}. When we focus on the $\phi_+$ phases, which we assume come form a single hemisphere, we find that at higher altitudes the signal is both wider (has a larger FWHM) and less blueshifted. The blueshift evolution with altitude only shows a marginal trend (2.3-$\sigma$), but the FWHM trend is robust. Finally, we find that as we probe deeper in the atmosphere, the signal from the beginning of the transit is less significant (SNR=5.1 for the strong-line bin and SNR=2.84 for the weak-line bin) and peaks lower in the atmosphere (height ratios between the two half decrease from 0.76 for the strong-line bin to 0.44 for the weak-line bin). 

We next compared the data and the observed trends to GCMs with different prescriptions for magnetic drag by injecting the models into the data and performing the same analysis on the models.
%We find that all of the models underpredict the strength of the signal (potentially due to lower Fe I abundance or temperatures) and the total wind speeds (models do not show enough blueshifting), but show many consistent trends with the data. 
All of the models show the trend of larger FWHMs at higher altitudes and a stronger blueshifted signal at lower altitudes for the $\phi_+$ phase bin, as is seen in the data. We explain both of these trends as being due to the fact that at lower altitudes the models show smaller absolute velocities, but a more ordered velocity pattern due to the presence of a jet. At higher altitudes there are larger absolute velocities, but in a less ordered pattern, leading to broader CCFs with a smaller average blueshift. However, the models show key differences and we are able to rule out the uniform drag model due to the fact that the FWHMs for that model are larger at $\phi_+$ phases and the signal is stronger during the first half of the transit. As both of these trends are clearly ruled out by the data, we are able to conclude that applying a spatially uniform drag prescription via a single drag coefficient cannot reproduce the data. The remaining two models both adequately reproduce the RV shifts, FWHMs, and height ratios seen in the data, but we argue that while the 3G model does not predict the degree of blueshifting as well as the 0G model, it better follows the \textit{trends} seen in height ratio and FWHM. As all GCM models are unable to reproduce the degree of blueshifting \citep{Wardenier2021, savel2022}, and the 3G model also fit low resolution Spitzer phase curves better \citep{Beltz2022b}, we take this as tentative evidence of magnetic field effects shaping the environment of WASP-76b. While more work is needed to explore if other field strengths, physical processes, or more complex radiative transfer routines  can provide a better fit to all of the data, it is promising that a single GCM model can adequately fit both the low-resolution Spitzer phase curve and the high-resolution optical transmission spectra, as previous modeling efforts found that to fit key features in the data they required significantly different drag prescriptions \citep[e.g., ][]{Wardenier2021, savel2022, May2021}.

%This problem of underpredicted wind speeds has been present in all modeling work regarding WASP-76b up to this point, and previous works have had to invoke an added eccentricity or artificially removing Fe I to achieve a more blueshifted signal \citep{Wardenier2021, savel2022}. 

%The remaining two models, the 3~G and 0~G cases, follow the general trend in RV shifts and both models show larger blueshifts during the second half of the transit, as well as greater blueshifts deeper in the atmosphere at $\phi_+$ phases. However, we find that the 0G model better matches the degree of blueshifting (does not underestimate the blueshifts as much) and has a larger difference between the RV shift at the beginning and end of the transit, as seen in the data. On the other hand, the 3G model more closely matches the trends in height ratio seen in the data. We can conclude that the active magnetic model (3G) is preferred over the uniform drag model, but the quality of the data does not show a strong preference between the purely hydrodynamical case (0G) and our active drag. As the 3G model also fit low resolution Spitzer phase curves better, this is potential evidence 

This study is a first step towards exploring exoplanets along new dimensions. As detection signal strengths increase, it becomes possible to better resolve the atmosphere by breaking the signal up. \citet{Ehrenreich2020}, and later \citet{Kesseli2022} and \citet{Seidel2023}, showed that by observing the shape and velocity evolution of planetary absorption lines during a transit, differences between the two terminators of a planet can be inferred. \citet{Gandhi2022} performed retrievals that successfully characterized WASP-76b's atmosphere at 4 discreet locations across the planetary surface by breaking the transit up both temporally (probing different regions as the planet rotates during transit) and in velocity space (separating the oppositely rotating eastern and western hemisphere). In this work, we present a novel method to probe the vertical structure of the planet's atmosphere by taking advantage of the fact that absorption lines of different opacity become optically thick at different altitudes. With the current limits on signal, we are unable to draw extensive conclusions or distinguish between a 3-G and 0-G model, but with future facilities such as high resolution spectrographs on ELTs, this will be possible. 

%We therefore we can conclude that the more complex magnetic model prescriptions are preferred over the uniform drag model, but the quality of the data at this point are not enough to distinguish between these two models or make any statements about a magnetic field on WASP-76b. 

%The 3~G model more closely matches the trends in height ratio shown by the data, making it slightly preferred. This is significant, as it means the 3~G is prefferred over the 0~G and uniform models in both high-resolution (presented here) and low resolution phase curves. 

\acknowledgements

We would like to thank Adam Langeveld for useful discussions that prompted inclusion of the $K_p$ vs. $V_{sys}$ diagrams. This publication makes use of The Data \& Analysis Center for Exoplanets (DACE), which is a facility based at the University of Geneva (CH) dedicated to extrasolar planets data visualisation, exchange and analysis. DACE is a platform of the Swiss National Centre of Competence in Research (NCCR) PlanetS, federating the Swiss expertise in Exoplanet research. The DACE platform is available at https://dace.unige.ch. Many of the calculations in this paper made use of the Great Lakes High Performance Computing Cluster maintained by the University of Michigan.

\bibliographystyle{aasjournal}
\bibliography{bib.bib}

\appendix
\section{Determining Uncertainties}
\label{A:unc}

As our conclusions are drawn from the measured radial velocities and their uncertainties, ensuring the accuracy of our values and uncertainties is vital for the interpretation of the data and in turn the robustness of our conclusions. In this section we try to calculate errorbars in a few different ways in order to ensure that they give consistent results and are truly representative of the data. We also compare our uncertainties to other publications of different data sets and discuss how the reduction and analysis steps affect the measured velocities to ensure that there are no hidden factors that are not represented in the errorbars. 

In the simplest method, we used \texttt{lmfit} (a python package for curve fitting) to fit Gaussians to the CCFs. As the data are noisy and we do not know the underlying functional form of the data (e.g., Gaussian, double-peaked, etc.), some fitting or smoothing needs to be done to obtain measured values and Gaussians provides the best solution given the data quality and information. \texttt{Lmfit} reports uncertainties in the parameters it fits, one of which is the center of the Gaussian. These uncertainties come from the covariance matrices. Using the CCF from the strong line bin during the second half of the transit as an example (black line in top panel of Figure \ref{f:1DCCF_3bins_split}), we obtain a best-fit value of $-9.565 \pm 0.479$ km s$^{-1}$ from \texttt{lmfit}. Using the same fitted Gaussians as above, but instead of simply assuming the uncertainties given by \texttt{lmfit}, we calculate the standard deviation of the fitted Gaussians and divide by the signal to noise ratio of the CCF \citep[as was done in ][]{Kesseli2021}. This method gives an uncertainty of 0.348 km s$^{-1}$ for the same example as above. Finally, we also calculate uncertainties by recording the velocity shift of each individual CCF in time and then taking the standard deviation of that measurement and dividing by the square root of the number of measurements. This method was also used to derive alternative uncertainty values in \citet{Kesseli2021}, and is similar to one of the methods used to derive radial velocity uncertainties in \citet{Wardenier2024}. This final method assumes that the true radial velocity is same at every measurement (at least within a small phase bin), and therefore that any change in velocity is uncertainty in the measurement and not physical. This is not the case for at least the beginning of the transit, but it appears to be a reasonable assumption for the end of the transit (looking at Figure \ref{f:2DCCFs}, phases from 0.02 to 0.04 have a vertical residual pattern). Using this approach, again for the same example as the above two, we are able to get 4 separate measurements of the radial velocity and find an uncertainty of 0.341 km s$^{-1}$. The fact that the three methods explored here return similar uncertainties lends credence to our uncertainty results. Throughout the paper, we quote the uncertainties as those from \texttt{lmfit} as these seem to be the most conservative and are the simplest to derive. 

Next, we compare our uncertainty values to other recent papers to ensure that our values are on the same order as these.  \citet{Borsa2021} use a single ESPRESSO transit of WASP-121b and report a radial velocity uncertainty of 0.16 and 0.28 km s$^{-1}$ in the first and second half of the transit, respectively, when the CCF is split in two phase bins in a similar manner as we have done. Our uncertainties, which are on the order of 0.4 km s$^{-1}$, seem reasonable considering we have two transits, but we split our CCFs into 3 line-strength bins before we split by the two phase bins. 

Finally, there may be data reduction or analysis steps that cause small changes in the measured radial velocities. This uncertainty in the radial velocity would not be taken into account in the above calculations, and so we explore how some of our choices affect the measured radial velocities. We obtain consistent measurements for the radial velocity of the Fe I signal with \citet{Kesseli2022} ($\phi_-=-3.89$km s$^{-1}$, $\phi_+=-10.26$km s$^{-1}$), and  \citet{Ehrenreich2020}, as reported by \citet{Wardenier2021} (averaging $\phi_-$ phases $\sim-3.5$km s$^{-1}$, and $\phi_+\sim-10.5$km s$^{-1}$), using a completely different method to either paper. In our own analysis, we do not perform any cleaning steps during our analysis process that significantly alters the data, unlike similar analyses in the NIR, which use PCA to clean the data. Indeed, \citet{Wardenier2024} calculate uncertainties in their radial velocity measurements by using different numbers of PCA iterations and taking the resulting radial velocities as different ''measurements" to obtain errorbars, but as our analysis does not perform any drastic cleaning steps, an approach like this is not possible. We did test a few of our data processing choices (i.e., masking bad pixels, etc.) and found that the resulting measured radial velocity differences were smaller than the uncertainties we quote in the paper. The only choice that made a significant difference in the measured radial velocities was the choice of how many lines to include in each mask, but as long as the same choice is used during the GCM model analysis and throughout the analysis, the comparisons and conclusions will be accurate. We therefore do not think that the analysis itself imparts significant uncertainties into the measured radial velocities, and that both the values and uncertainties accurately represent the data. 

The data themselves are noisy because we have split the signal into so many different parts and due to the inherent increased noise when using the binary mask method as opposed to a traditional CCF method, but we note that every split CCF has a SNR$>$4.5, which is usually considered a reliable detection for high resolution observations. Because of this, we believe that the data are of high enough quality to measure reliable RVs. As data quality improves with the coming of ELTs, more robust measurements will be possible that rely on fewer underlying assumptions (e.g., Gaussian shapes).

\section{Trends with altitude in Measured K$_p$}
\label{A:KpVsys}

\begin{figure}
\begin{center}
\includegraphics[width=0.48\linewidth]{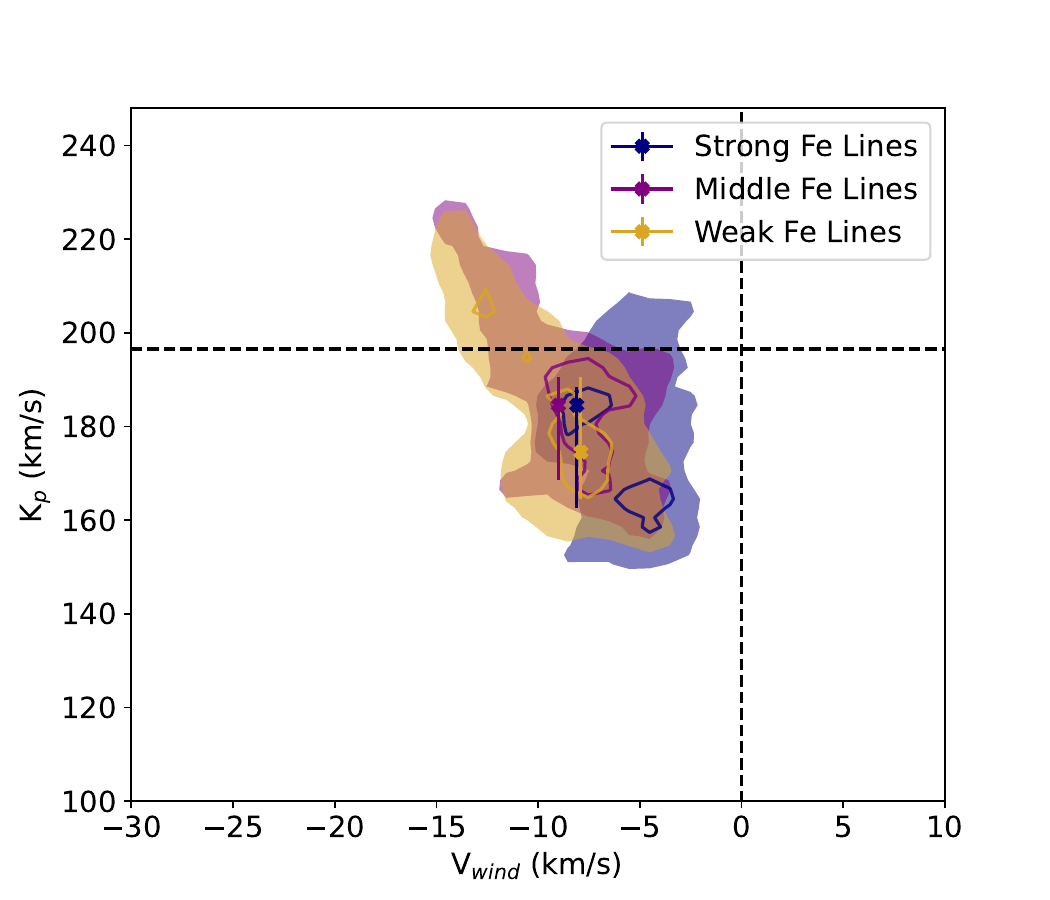}
\includegraphics[width=0.48\linewidth]{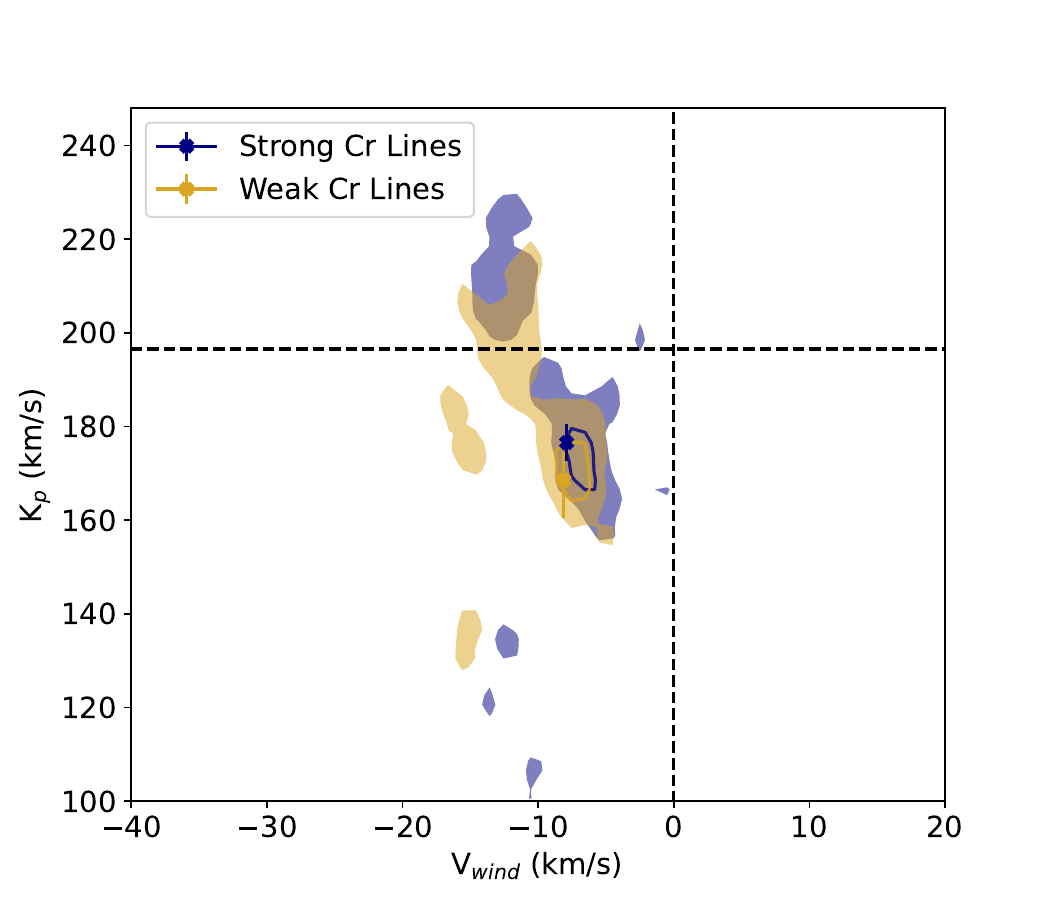}
\includegraphics[width=0.48\linewidth]{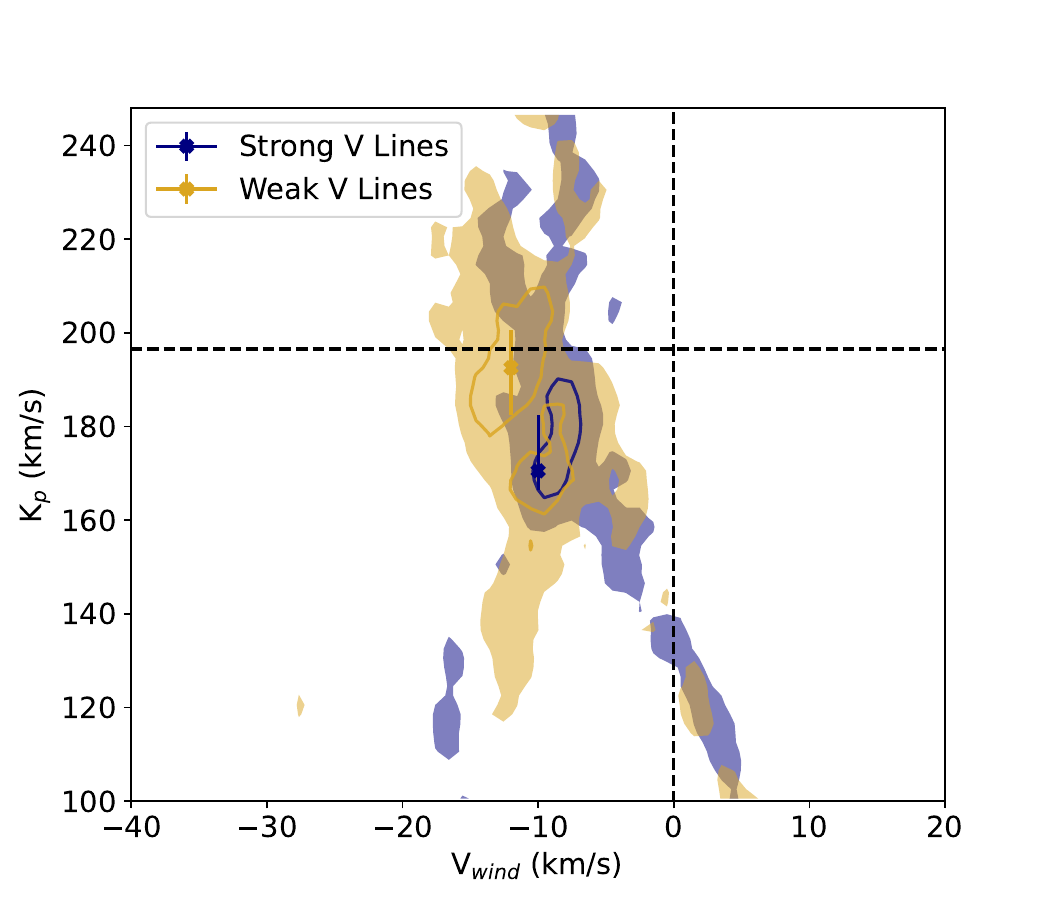}
\includegraphics[width=0.48\linewidth]{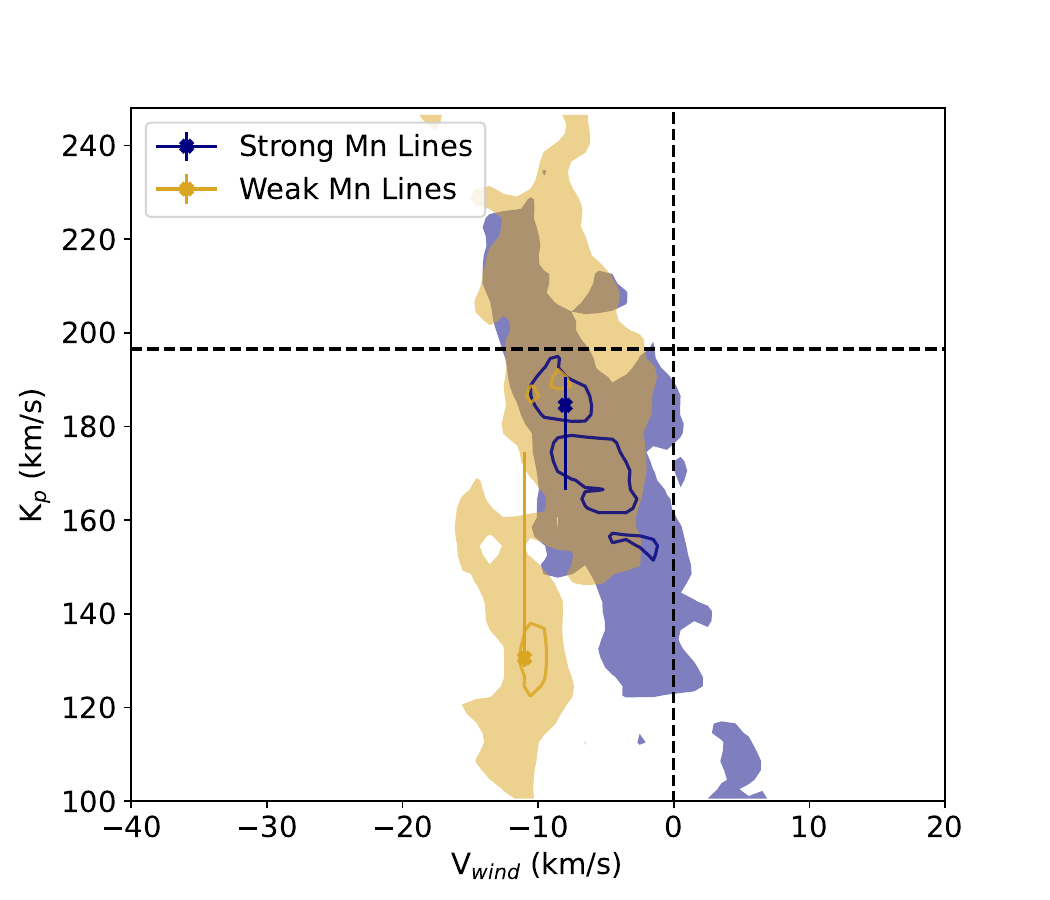}
\caption{\small $K_p$ versus $V_{wind}$ plots for Fe I (top left), Cr I (top right), V I (bottom left), and Mn I (bottom right). The different colors are overlays of SNR contours for our different altitude bins, and the crosses denote the peak $K_p$ and $V_{wind}$ value for each altitude bin. The semi-transparent filled contours denote where the SNR has dropped by 1.5 from the peak SNR value, while the line contours of the same color denote where the SNR has dropped by 0.5. In some cases, there are multiple regions of high SNR, which can be seen as multiple contours (see for example the two different navy blue lined contours for the strong Fe line bin). We have also used these contours to estimate uncertainties in the peak $K_p$, which are plotted as errorbars on the crosses. We find no clear trend in $K_p$ offset with altitude. }
\label{f:KpVsys}
\end{center}
\end{figure}

Motivated by recent work that find differing offsets for different chemical species in the standard $K_p$ vs $V_{sys}$ diagram and suggested this difference could be due to species residing in different layers (altitudes) of a planet's atmosphere \citep[e.g., ][]{Kesseli2022, Borsato2023, Brogi2023}, we tested to see whether our CCFs created with different altitude bins showed noticeable offsets in a $K_p$ vs $V_{sys}$ diagram. 

These diagrams are created in the standard method, by taking the 2D CCFs for each species and each altitude bin and shifting them to rest assuming different values of planet semi-amplitude ($K_p$) and different systemic velocities ($V_{sys}$). Once the 2D CCFs are shifted to rest, they are co-added in this rest frame and the SNR is calculated as described in Section \ref{s:cc}. In order to compare the positions of the peaks in these diagrams, we have overlayed the SNR contours for the different altitude bins of each individual species in a single plot (Figure \ref{f:KpVsys}). By comparing the peaks for each bin (marked with crosses) and their associated SNR contours, we do not see a clear pattern emerge of differing $K_p$ offsets at different altitudes. For the most part, the measured SNR peaks for a single species are consistent to within 1-$
\sigma$, and the strong line peak is not always offset from the weak line peak in the same direction (i.e. the strong line bin peak for V is at a higher $K_p$ value than the weak line bin, whereas for Cr and Mn it is the opposite). 

We see this lack of trend as potential indication that altitude is not the sole driver of differences in the measured peak $K_p$, and other processes, such as condensation or ionization, may lead to more significant differences in the measured $K_p$ offset. It is important to note, however, that the species studied here only range in altitude from about 1.01 to 1.10 $R_p$ ($\sim10^{-3} - 10^{-8}$ bar), and so species that primarily absorb in regions of the atmosphere outside of this range (e.g., H, He, etc.) may show significantly offset $K_p$ values that are indicative of their altitudes.

\end{document}